\documentclass[]{aa} 
\usepackage{graphicx}
\usepackage{txfonts}
\begin{document}
   \title{Molecular Gas in NUclei of GAlaxies (NUGA)}

   \subtitle{IV.~Gravitational Torques and AGN Feeding \thanks{Based on 
observations carried out with the IRAM Plateau de Bure Interferometer. 
IRAM is supported by INSU/CNRS (France), MPG (Germany) and IGN (Spain)}}

   \author{S.~Garc\'{\i}a-Burillo\inst{1}
          \and
	   F.~Combes\inst{2}
          \and
	   E.~Schinnerer\inst{3}
	  \and
	   F.~Boone\inst{4} 
	  \and 
	   L.~K.~Hunt\inst{5}}

   \offprints{S.~Garc\'{\i}a-Burillo}

   \institute{Observatorio Astron\'omico Nacional (OAN)-Observatorio de Madrid,
Alfonso XII, 3, 28014-Madrid, Spain \\
              \email{s.gburillo@oan.es}
         \and	 
             Observatoire de Paris, LERMA, 61 Av. de l'Observatoire, 75014-Paris,
France \\
             \email{francoise.combes@obspm.fr}	
	\and	
	 Max-Planck-Institut f\"ur Astronomie, K\"onigstuhl, 17, 69117-Heidelberg, Germany \\
             \email{schinner@mpia-hd.mpg.de}	     
	\and	
	    Max-Planck-Institut f\"ur Radioastronomie, Auf dem H\"ugel, 69, 53121-Bonn, Germany.\\
             \email{fboone@mpifr-bonn.mpg.de}
	\and
	     Istituto di Radioastronomia/CNR, Sez. Firenze, 
		Largo Enrico Fermi, 5, 50125-Firenze, Italy \\
             \email{hunt@arcetri.astro.it}
	    }

\date{Received February 17th, 2005; accepted ---, 2005}

   \abstract{We discuss the efficiency of stellar gravity torques as a
mechanism to account for the feeding of the central engines of four low luminosity Active
Galactic Nuclei (AGN): NGC\,4321 (HII nucleus/LINER), NGC\,4826 (HII nucleus/LINER), 
NGC\,4579 (LINER 1.9/Seyfert 1.9) and NGC\,6951 (Seyfert 2). These galaxies have been observed as part of the {\it
NUclei of GAlaxies}--(NUGA) CO project, aimed at the study of AGN fueling mechanisms.
Our calculations allow us to derive the characteristic time-scales for gas flows and
discuss whether torques from the stellar potentials are efficient enough to drain the gas
angular momentum in the inner 1~kpc of these galaxies. The stellar potentials are derived using
high-resolution near infrared (NIR) images and the averaged effective torques on the gas are
estimated using the high-resolution ($\sim$0.5''--2'') CO maps of the galaxies. Results
indicate paradoxically that feeding {\it should be} thwarted close to the AGNs: in the four cases
analyzed, gravity torques are mostly positive inside r$\sim$200~pc, resulting in no inflow on these scales. 
As a possible solution for the paradox, we speculate that the agent responsible for driving inflow to still 
smaller radii is transient and thus presently absent in the stellar potential. Alternatively, the gravity torque barrier associated
with the Inner Lindblad Resonance of the bars in these galaxies could be overcome by
other mechanisms that become competitive {\it in due time} against gravity torques. In particular,
we estimate on a case-by-case basis the efficiency of viscosity versus gravity
torques to drive AGN fueling. We find that viscosity can counteract moderate--to--low gravity
torques on the gas if it acts on a nuclear ring of high gas surface density contrast and $\sim$a few 100\,pc size.

We propose an evolutionary scenario in which gravity torques and viscosity act in concert to produce
recurrent episodes of activity during the typical lifetime of any galaxy. In this scenario the
recurrence of activity in galaxies is indirectly related to that of the bar instabilities although
the active phases are not necessarily coincident with the maximum strength of a single bar episode.
The general implications of these results for the current
understanding of fueling of low-luminosity AGN are discussed.

   }

   \maketitle
%

\section{Introduction}

The phenomenon of nuclear activity is understood to be a result of the feeding of supermassive black holes (SMBHs) in 
galactic nuclei. Observational evidence accumulated over the last decade indicates that SMBH exist 
in most galactic bulges (e.g., Kormendy \& Richstone~\cite{kor95}; Magorrian et al.~\cite{mag98}; 
Ferrarese \& Merritt~\cite{fer00}; Gebhardt et al.~\cite{geb00}). Among all these massive black holes,
very few are highly active. AGN are found in 10$\%$ of the local galaxies (Ho et al.~\cite{ho97}); however, this percentage 
is increased up to $\sim$44$\%$ if LINERs are taken into account. One of the present 
challenges is to understand how AGN can be fed during their lifetime. In the feeding problem 
 the gas supply must come from the whole disk of the host galaxy at large distances 
compared to the radius of gravitational influence of the central engine. Therefore it is expected that a hierarchy 
of mechanisms combine to drive virtually all the gas from the large 
$\sim$kpc scales down to the inner $\sim$pc scales. The different spatial scales involved  
suggest that the various mechanisms at work have very different time scales (Shlosman et 
al.~\cite{shl89,shl90}; Combes~\cite{com01,com03}; Jogee~\cite{jog04}).  Recent observational and 
theoretical evidence indicates that the AGN lifetimes may be as short as $\sim$a few 10$^{7}$--10$^{8}$yr (Ho et al.~\cite{ho03}; Martini~\cite{mar04}; Wada~\cite{wad04}; Merloni~\cite{mer04}). Moreover, Wada~(\cite{wad04}) finds evidence that 
mass accretion may not be constant even during the nominal duty cycle of 10$^{8}$yr, but composed of several shorter episodes
with a duration of 10$^{4-5}$yr. This time-scale conspiracy could explain the lack of success of observers in finding any  
correlation between the presence of $\sim$kpc scale non-axisymmetric perturbations (e.g.,
large-scale bars and interactions) and the onset of activity in galaxies, except for very high
luminosity objects(QSOs) (Moles et al.~\cite{mol95}; Mulchaey \& Regan~\cite{mul97}; Knapen et
al.~\cite{kna00}; Krongold et al.~\cite{kro01}; Schmitt~\cite{sch01}). On these spatial scales, Hunt \&
Malkan~(\cite{hun99}) have found a significant correlation between the detection
rate of outer rings and the onset of activity.

\begin{table*}
\begin{tabular}{|l|c|c|c|c|c|c|c|c|c|}
\hline\hline
\multicolumn{1}{|l|}{Galaxy} &
\multicolumn{1}{c|}{$PA$($^{\circ}$)} &
\multicolumn{1}{c|}{$i$($^{\circ}$)} &
\multicolumn{1}{c|}{$D$(Mpc)} &
\multicolumn{1}{c|}{scale(pc/$^{\prime\prime}$)} & 
\multicolumn{1}{c|}{RA$_{2000}$} & 
\multicolumn{1}{c|}{Dec$_{2000}$} & 
\multicolumn{1}{c|}{v$_{sys}^{LSR}$(km~s$^{-1}$)} & 
\multicolumn{1}{c|}{CO(1--0)($^{\prime\prime} \times ^{\prime\prime}$)} &
\multicolumn{1}{c|}{CO(2--1)($^{\prime\prime} \times ^{\prime\prime}$)} \\
\hline\hline
{\bf NGC~4321} & 153 & 32 & 16.8 & 83 & 12$^h$22$^m$54.91$^s$ &	15$^o$49$^{\prime}$19.9$^{\prime\prime}$ & 1573$\pm$10  & 2.2$\times$1.2 & -- \\
{\bf NGC~4826} & 112 & 54 & 4.1 & 20 & 	 12$^h$56$^m$43.63$^s$	& 21$^o$40$^{\prime}$59.1$^{\prime\prime}$	& 413$\pm$10  & 2.5$\times$1.8 & 0.7$\times$0.5 \\
{\bf NGC~4579} & 95 & 36 & 19.8 & 97 & 	12$^h$37$^m$43.52$^s$	& 11$^o$49$^{\prime}$05.5$^{\prime\prime}$	& 1469$\pm$10	& 2.0$\times$1.3 & 1.0$\times$0.6 \\
{\bf NGC~6951} & 134 & 42 & 19 & 93 & 	20$^h$37$^m$14.12$^s$	& 66$^o$06$^{\prime}$20.0$^{\prime\prime}$	& 1441$\pm$10	& 1.4$\times$1.1 & 0.6$\times$0.5 \\
\hline\hline
\end{tabular}
\caption{We list the position angle ($PA$), inclination ($i$), distance (D), spatial scale, coordinates of the dynamical center derived from CO and/or from the locus of the radio continuum source [(RA$_{2000}$, Dec$_{2000}$); accurate to $\sim$0.5$^{\prime\prime}$ on average] and systemic velocity(v$_{sys}^{LSR}$) derived from CO for NGC~4321, NGC~4826, NGC~4579 and NGC~6951 used in this paper. The spatial resolutions of the CO observations are listed in columns 9--10.}
\label{parameters}
\end{table*}

The search for a {\it universal} feeding mechanism has been pursued by looking for morphological features in the 
central kpc of nearby AGN with high spatial resolution ($\sim$a few 100~pc), though with limited success. Nuclear 
stellar bars seem to be as common in AGN as in non-AGN (Regan \& Mulchaey~\cite{reg99}; Laine et 
al.~\cite{lai02}). Furthermore, a similar scenario holds for nuclear spirals: while Martini \& 
Pogge~(\cite{mar99}) and Pogge \& Martini~(\cite{pog02}) initially found a high frequency of dusty spirals in the 
HST enhanced color images of Seyfert nuclei of their sample, more recently, Martini et al.~(\cite{mar03}) have 
shown on a firmer statistical basis that these nuclear features are not preferentially found in AGN.
Similarly there is only weak statistical evidence that nuclear rings are more frequently found in
Seyferts (Knapen~\cite{kna05}). Hunt \& Malkan~(\cite{huntmalkan04}) found evidence for an excess in type 2 Seyferts of 
kpc-scale twisted isophotes, though the cause of such twists is unclear.

On the modeling front, significant progress has been made in recent years on the study of the feeding 
efficiency of different types of gravitational instabilities: nested bars (e.g., Shlosman et al.~\cite{shl89}; 
Friedli \& Martinet~\cite{fri93}; Maciejewski \& Sparke~\cite{mac00}; Englmaier et al.~\cite{eng04}), gas spiral 
waves (e.g., Englmaier \& Shlosman~\cite{eng00}; Maciejewski et al.~\cite{mac02}; Maciejewski~\cite{mac04a, 
mac04b}), $m=1$ perturbations (e.g., Shu et al.~\cite{shu90}; Junqueira \& Combes~\cite{jun96}; 
Garc\'{\i}a-Burillo et al.~\cite{gb00}) and nuclear warps (e.g., Schinnerer et al.~\cite{sch00}).
However, the lack of high quality multi-wavelength observational constraints on the different models has thus
far made the choice of a single optimal scenario rather difficult.

The study of interstellar gas in AGN is essential to understand the phenomenon of nuclear activity
in galaxies and its possible link to circumnuclear star formation. As most of the neutral gas in
galactic nuclei is in the molecular phase, CO lines are best suited to 
undertake high-resolution mapping of AGN hosts, with interferometer resolution of $<$100~pc, i.e.,
the scales on which {\it secondary} modes embedded in kpc-scale perturbations are expected to take over. 
CO lines better trace the total gas column densities than dust extinction probes obtained from HST NIR/optical color images. Most importantly, CO maps provide the gas kinematics (velocity fields and velocity dispersions). This
information is essential to characterize gravitational instabilities and to constrain the
models.  The NUclei of GAlaxies--NUGA--project, fully described by Garc\'{\i}a-Burillo et
al.~(\cite{gb03,paperI}), is the first high-resolution
($\sim$0.5$^{\prime\prime}$--1$^{\prime\prime}$) CO survey of 12 low luminosity AGN (LLAGN)
including the full sequence of activity types (Seyferts, LINERs and transition objects from HII to LINER). 
In the case of LLAGN, the required mass accretion rates derived from the typical bolometric luminosities of these objects range from 10$^{-2}$ to 10$^{-5}$M$_{\odot}$yr$^{-1}$ (from Seyferts to LINERs; e.g., see compilation by Jogee~\cite{jog04}). Observations, carried out with the IRAM Plateau de Bure Interferometer (PdBI), have been completed early 2004. NUGA surpasses in both spatial resolution and sensitivity ongoing surveys of nearby AGN conducted at OVRO (MAIN: Jogee et al.~\cite{jog01}) and at NRO (Kohno et al.~\cite{koh01}).

In this paper we focus on the study of gravitational torques in a subset of NUGA galaxies,
which span the range of the different activity classes within our sample: NGC\,4321 
(transition object: HII/LINER), NGC\,4826 (transition object: HII/LINER), 
NGC\,4579 (LINER 1.9/Seyfert 1.9) and NGC\,6951(Seyfert 2). Information on the stellar potentials, 
obtained through available HST and ground-based optical/NIR
images of the sample, is used to determine the gravitational torques exerted by the derived stellar
potentials on the gaseous disk. The efficiency with which gravitational torques  drain the angular
momentum of the gas depends first on the strength of the non-axisymmetric perturbations of the
potential (m$>$0) but, also, on the existence of significant phase shifts between the gas and the
stellar distributions. The estimate of these phase shifts necessarily requires the availability of
images of comparably high spatial resolution ($\leq$0.5$^{\prime\prime}$ in our case) showing the
distribution of the stars and the gas. In this paper we purposely neglect the role of gas self-gravity as a source of non-axisymmetry in the gravitational potential. Nuclear galaxy disks with a high gas surface density and a mostly axisymmetric stellar potential can be prone to develop
this kind of gas self-gravitating perturbation. We purposely defer the study of {\it pure} gas instabilities and their ability to drive gas inflow to a forthcoming publication.

 We describe in Sect.~\ref{obs} the observations used, including high-resolution CO maps 
and NIR images of NGC\,4321, NGC\,4826, NGC\,4579 and NGC\,6951. 
Sect.~\ref{obs-feeding} interprets them in terms of AGN feeding.
Sect.~\ref{grav} computes from NIR images the gravitational potentials and forces,
and deduces from the CO maps the effective torques applied to the gas. From these torques, it is
possible to derive time-scales for gas flows and discuss whether gravity torques alone are
efficient enough to feed the AGNs. The general implications of these results for the current
understanding of AGN feeding are presented in Sect.~\ref{summary}.

\section{Observations\label{obs}}
\subsection{CO NUGA observations \label{obsco}}

 \begin{figure*}[bth!]
   \centering
   \includegraphics[width=18cm]{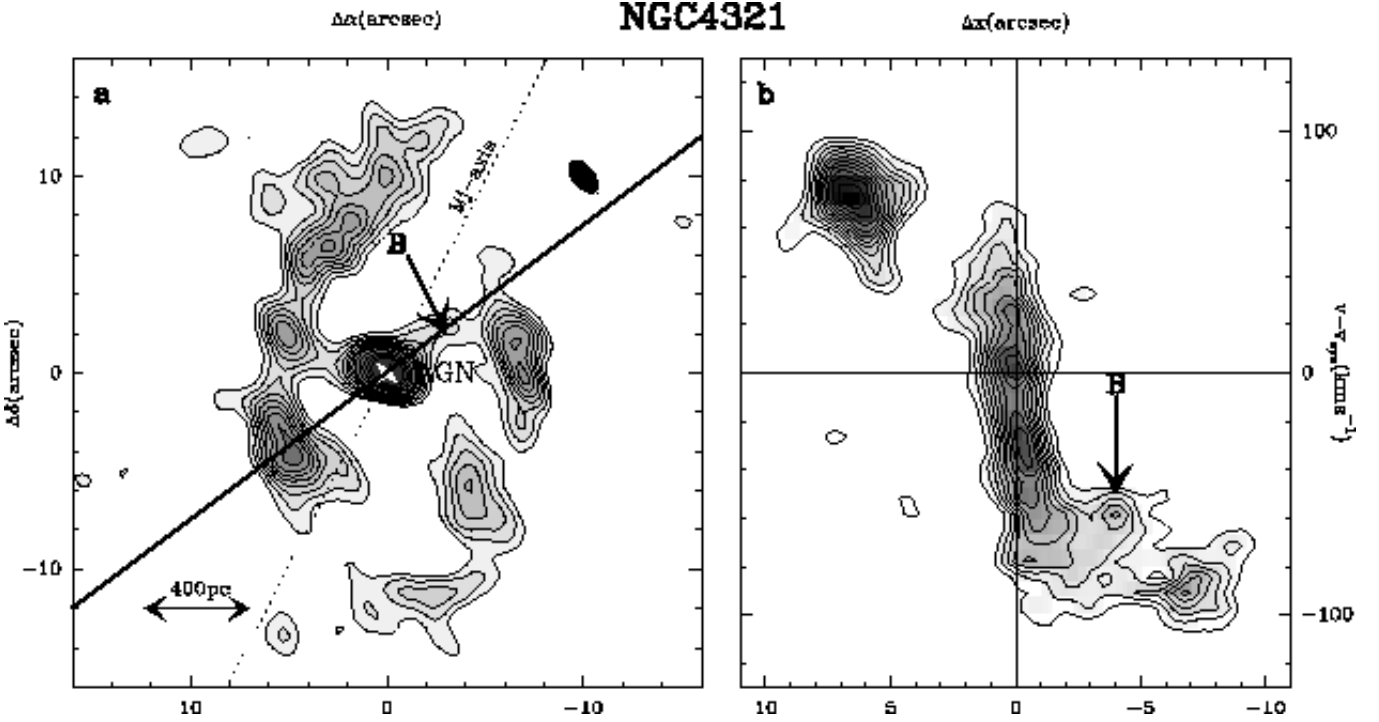}
   \caption{
   {\bf a)}~The $^{12}$CO(1--0) integrated intensity map obtained with the PdBI (contour levels from 1.5 to 5 
in steps of 0.5~Jy~km~s$^{-1}$~beam$^{-1}$ and from 5 to 10 in steps of 1~Jy~km~s$^{-1}$~beam$^{-1}$) observed in the nucleus of NGC\,4321. The 
filled ellipse at the top right corner represents the CO beam size. The central r$\sim$1~kpc molecular disk of 
NGC\,4321 consists of two nuclear spiral arms connected (e.g., {\bf B} component) to a marginally resolved 
r$\sim$150~pc disk of $\sim$10$^{8}$M$_{\odot}$ of molecular gas centered on the AGN locus (highlighted by the star marker).  ($\Delta\alpha$,~$\Delta\delta$)--offsets are with respect to the location of the AGN [(RA$_{2000}$, Dec$_{2000}$)=(12$^h$22$^m$54.91$^s$, 15$^d$49$^m$19.9$^s$)]. This position coincides within the errors with a secondary radio continuum maximum measured at 6~cm by Weiler et al.~(\cite{wei81}) and with a peak in the (J,H,K)-2MASS image of the galaxy (Jarrett et al.~\cite{jar03}).  Fig.~1{\bf b} illustrates the kinematics of molecular gas along the strip at PA=127$^{\circ}$ (thick line in Fig.~1{\bf a}). Levels go from 0.015 to 0.14 in steps of  0.0125~Jy~beam$^{-1}$. Velocities are relative to the systemic velocity (v$_{sys}^{LSR}$=1573~km~s$^{-1}$; determined from CO kinematics by Garc\'{\i}a-Burillo et al.~\cite{gb98}) and $\Delta$x offsets are relative to the AGN. The major axis orientation is shown by the dashed line. Gas kinematics over the emission bridge {\bf B} depart from circular rotation.
}
         \label{fig:feeding-4321}
   \end{figure*}


\begin{figure*}[bth!]
   \centering
   \includegraphics[width=18cm]{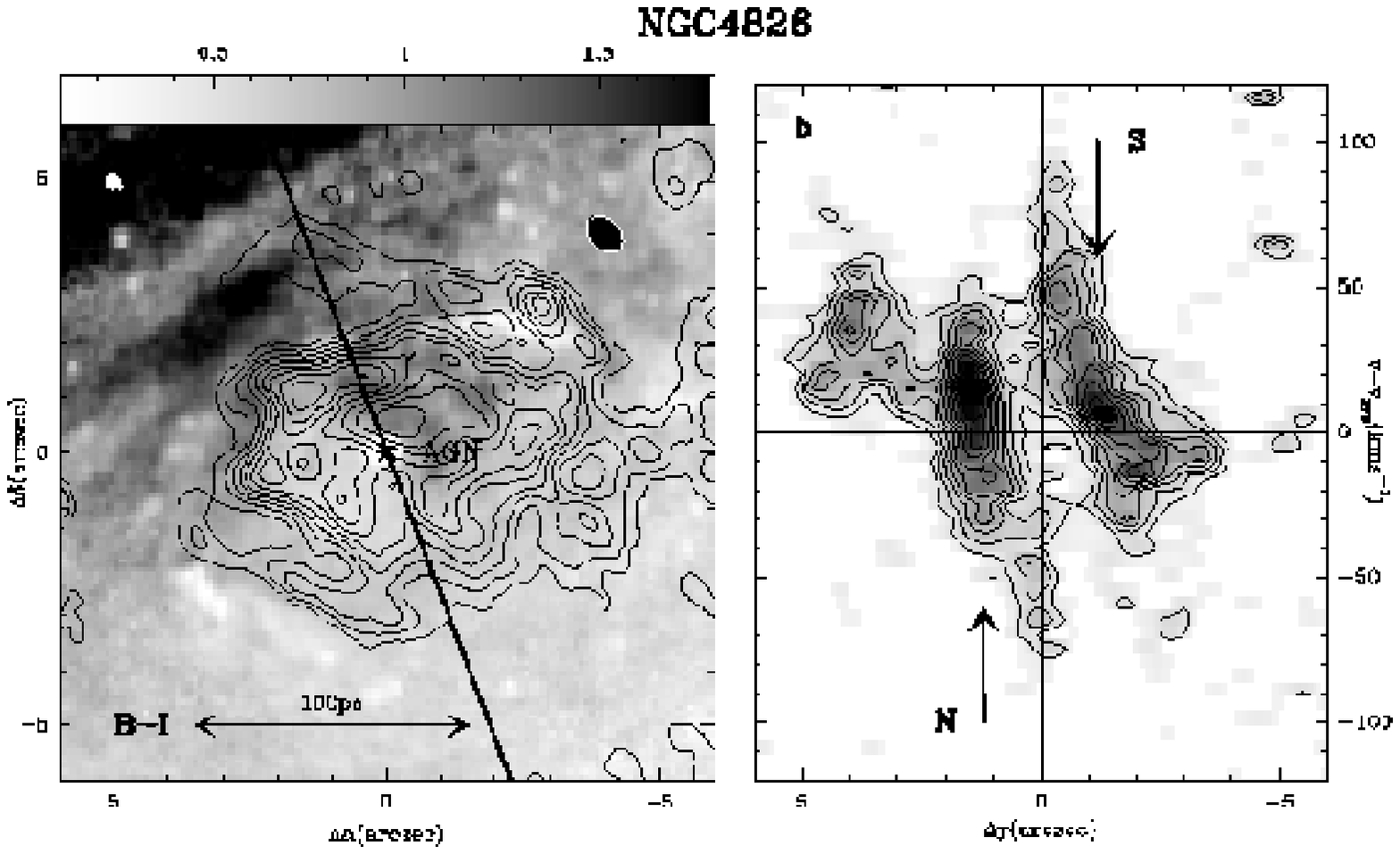}
   \caption{
   {\bf a)}~The $^{12}$CO(2--1) integrated intensity map obtained with the PdBI
(contour levels from 2.8 to 10 in steps of 0.9Jy~km~s$^{-1}$~beam$^{-1}$) is overlaid on the B-I color image
from HST (grey scale) observed in the nucleus of NGC\,4826. The filled ellipse
at the top right corner represents the CO beam size. ($\Delta\alpha$,~$\Delta\delta$)--offsets are with respect to the location of the AGN (marked by the star): (RA$_{2000}$, Dec$_{2000}$)=(12$^h$56$^m$43.63$^s$, 21$^d$40$^m$59.1$^s$). The AGN locus is identified by a blue point-like source in the B--I map; it also coincides with a non-thermal radio continuum peak measured at 6~cm by Turner \& Ho~(\cite{tur94}). The 80\,pc radius circumnuclear disk (CND) of NGC\,4826, with $\sim$3$\times$10$^{7}$M$_{\odot}$ of  molecular gas, shows a lopsided ringed disk morphology; the disk is off-center with respect to the AGN. {\bf b)} The kinematics of the gas, here displayed along the
minor-axis p-v plot (thick line in Fig.~2{\bf a}), are suggestive of strong streaming motions at the crossing of the
ring edges ({\bf N,~S}).  Levels go from  0.025 to 0.19 in steps of 0.015~Jy~beam$^{-1}$. Velocities are relative to the systemic velocity (v$_{sys}^{LSR}$=413~km~s$^{-1}$; determined from CO kinematics by Garc\'{\i}a-Burillo et al.~\cite{paperI}) and $\Delta$y offsets are relative to the AGN.
} 
\label{fig:feeding-4826} 
\end{figure*}


 \begin{figure*}[bth!]
   \centering
   \includegraphics[width=18cm]{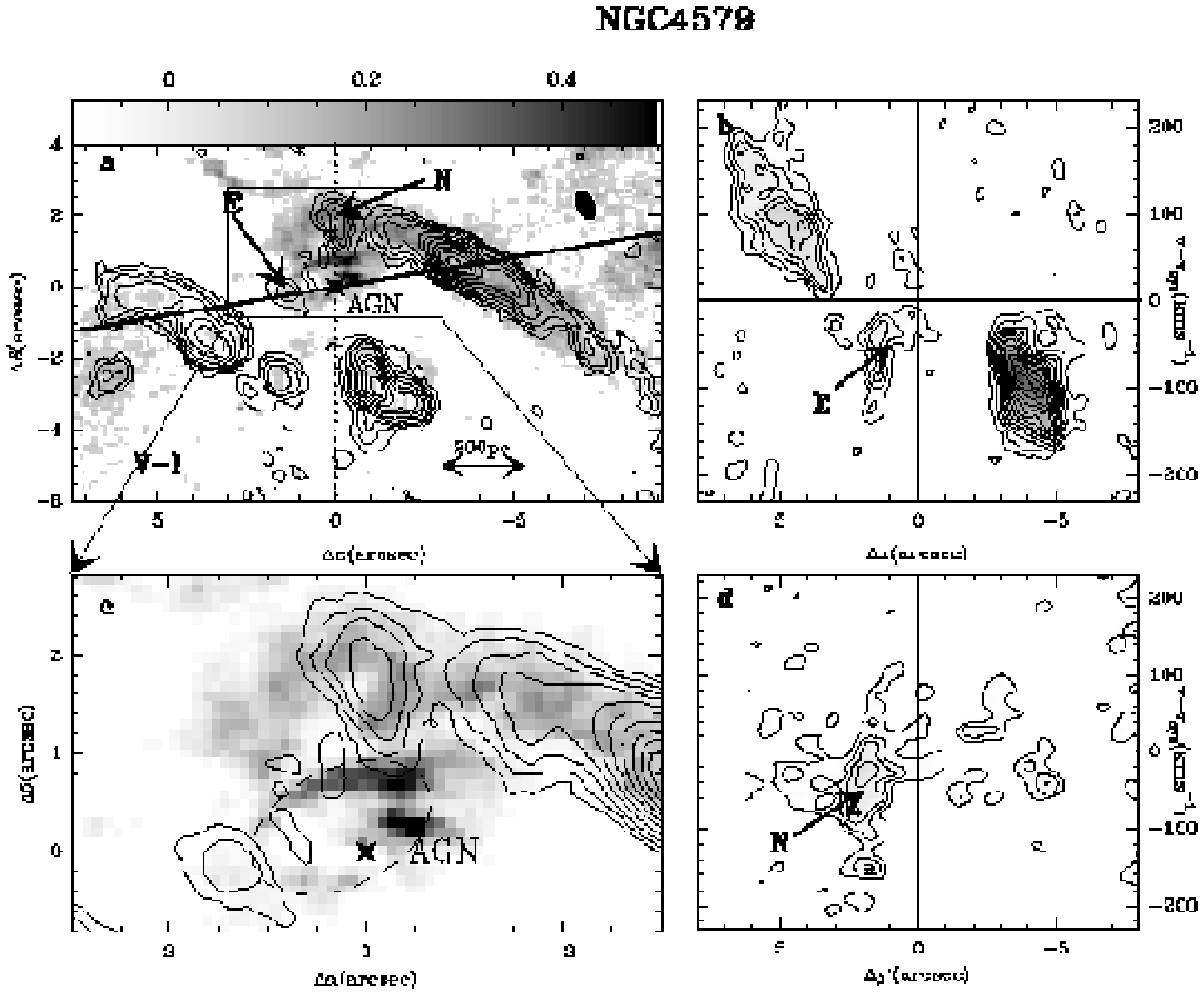}
   \caption{
   {\bf a)}~The $^{12}$CO(2--1) integrated intensity map obtained with the PdBI (contour levels from 0.4, 0.7, 1.3, 2.0 to 11 in steps of 
0.9Jy~km~s$^{-1}$~beam$^{-1}$) is overlaid on the V-I color image from 
HST (grey scale) observed in the nucleus of NGC\,4579. The filled ellipse at the top right corner represents the CO beam size.  ($\Delta\alpha$,~$\Delta\delta$)--offsets are with respect to the location of the AGN (marked by the star): (RA$_{2000}$, Dec$_{2000}$)=(12$^h$37$^m$43.52$^s$, 11$^d$49$^m$05.5$^s$). The AGN locus is identified by a point-like continuum source detected at 3~mm and 1~mm by Garc\'{\i}a-Burillo et al.~2005 (in prep). Two spiral arcs 
concentrate the bulk of the $\sim$3.2$\times$10$^{8}$M$_{\odot}$ molecular gas mass in the central 1~kpc of the 
galaxy. A close-up view of the inner 200~pc region (shown in {\bf c}), shows a central ringed disk
(highlighted by the dashed ellipse) with the AGN lying on its southwestern edge. A gas clump of
$\sim$10$^{6}$M$_{\odot}$ (denoted as {\bf E}) delimits the disk to the East. {\bf b)} The kinematics along the
major axis 
(thick line in Fig.~3{\bf a}) reveal highly non-circular motions related to the {\bf E} clump. Levels go from 0.006
to 0.09 in steps of 0.006~Jy~beam$^{-1}$. Velocities are relative to the systemic velocity
(v$_{sys}^{LSR}$=1469~km~s$^{-1}$; determined from CO kinematics by Garc\'{\i}a-Burillo et al.~2005, in prep.) and
$\Delta$x offsets are relative to the AGN. {\bf d)} Same as {\bf b)} but here along the declination axis (dashed
thick line in Fig.~3{\bf a}) with levels going from 0.006 to 0.024 in steps of 0.006~Jy~beam$^{-1}$. Highly
non-circular motions are related to the {\bf N} clump.  
   }
         \label{fig:feeding-4579}
   \end{figure*}


 \begin{figure*}[bth!]
   \centering
   \includegraphics[width=18cm]{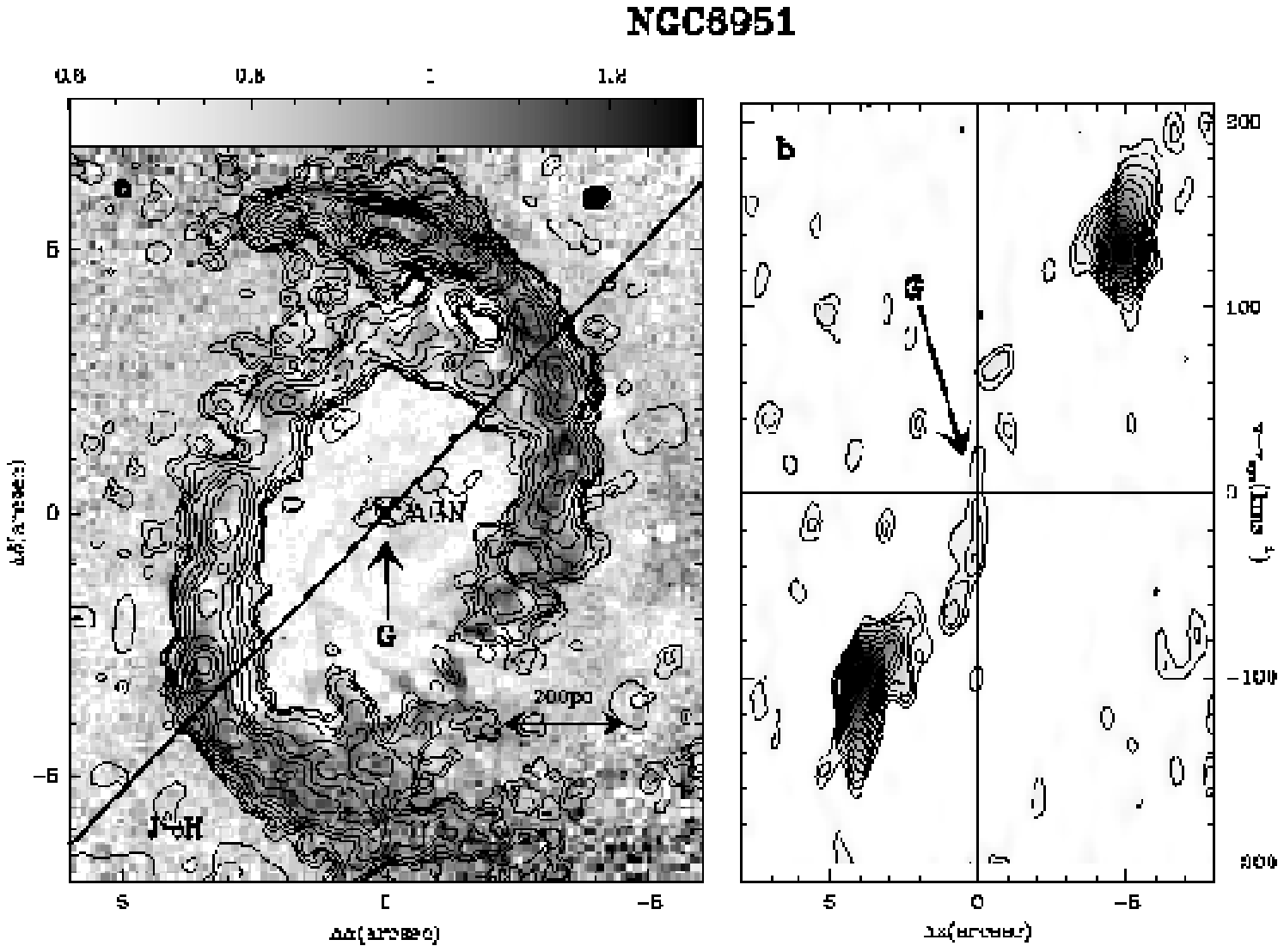}
   \caption{
   {\bf a)}~The $^{12}$CO(2--1) integrated intensity map obtained with the PdBI (contour levels from 0.15, 0.25, 0.40, 0.70 to 6.2
in steps of 0.5Jy~km~s$^{-1}$~beam$^{-1}$) is overlaid on the
J-H color image from HST (grey scale) observed in the nucleus of NGC\,6951. The filled ellipse at the top right corner represents the CO
beam size. ($\Delta\alpha$,~$\Delta\delta$)--offsets are with respect to the location of the AGN (marked by the star): (RA$_{2000}$, Dec$_{2000}$)=(20$^h$37$^m$14.12$^s$, 66$^d$06$^m$20.0$^s$). The position of the AGN is given by the point-like radio continuum source measured at 6~cm and 20~cm by Ho \& Ulvestad~(\cite{ho01}). The molecular gas distribution in the central 1\,kpc shows two highly contrasted nuclear spiral arms containing 3$\times$10$^{8}$M$_{\odot}$ which are presently feeding a circumnuclear starburst. A
compact molecular complex (denoted as {\bf G}) of $\sim$a few 10$^{6}$M$_{\odot}$ is detected on the
AGN. Furthermore, we have tentatively  detected a northern molecular gas component linking {\bf G} with the N spiral arm (i.e., the spiral running North from  West) which could be related to the filamentary dusty spiral seen in the J--H color HST image (see also Fig.~4b). {\bf b)} The kinematics along the major axis (thick line in Fig.~4{\bf a}) are compatible with circular motions for the gas near the AGN. Levels go from 0.008, 0.0012, 0.016 to 0.076 in steps of 0.006~Jy~beam$^{-1}$. Velocities are relative to the systemic velocity (v$_{sys}^{LSR}$=1441~km~s$^{-1}$; determined from CO kinematics by Schinnerer et al.~2005, in prep.) and $\Delta$x offsets are relative to the AGN.
}
 \label{fig:feeding-6951}
   \end{figure*}


Observations of the circumnuclear disks of NGC\,4826, NGC\,4579 and NGC\,6951 were carried out as
part of the NUGA survey with the PdBI between December
2000 and March 2003. We used the ABCD set of configurations of the array (Guilloteau et
al.~\cite{gui92}). This assures high spatial resolution ($<$1$^{\prime\prime}$ at the highest
frequency) but also an optimum sensitivity to all spatial frequencies in the maps. A previous set of
observations of NGC\,4826, using data taken with the BCD configurations, has been discussed in
Garc\'{\i}a-Burillo et al. (\cite{paperI}) (hereafter called paper I). We have observed
simultaneously the J=1--0 and J=2--1 lines of $^{12}$CO in single fields. Table~\ref{parameters} 
lists the spatial resolution of these CO observations as well as other source related parameters.
The primary beam size is 42$^{\prime\prime}$ (21$^{\prime\prime}$) in all the 1--0 (2--1) line
observations. As the bulk of the relevant nuclear disk emission arises well inside the central 15$^{\prime\prime}$ in the four galaxies discussed in this paper, observations have not been corrected for primary beam attenuation. 
 During the observations the spectral correlator was split in two halves centered at
the transition rest frequencies corrected for the assumed recession velocities. The correlator
configuration covers a bandwidth of 580\,MHz for each line, using four 160\,MHz-wide units; this is
equivalent to 1510\,km~s$^{-1}$(755\,km~s$^{-1}$) at 115\,GHz (230\,GHz). Visibilities were obtained
using on-source integration times of 20 minutes framed by short ($\sim 2$\,min) phase and amplitude
calibrations on nearby quasars. The absolute flux scale in our maps was derived to a 10\%
accuracy based on the observations of primary calibrators whose fluxes were determined from a combined set of measurements obtained at the 30m telescope and the PdBI array. Image reconstruction was done using standard IRAM/GAG software (Guilloteau \& Lucas~\cite{gui00}). In this work we use prior observations of the 1--0 line of $^{12}$CO of
NGC\,4321 made using the BCD configurations of the PdBI and previously published by
Garc\'{\i}a-Burillo et al.~(\cite{gb98}) (see this paper and Table~\ref{parameters} for details).

\begin{figure}[tbh!]
   \centering
   \includegraphics[width=8.5cm]{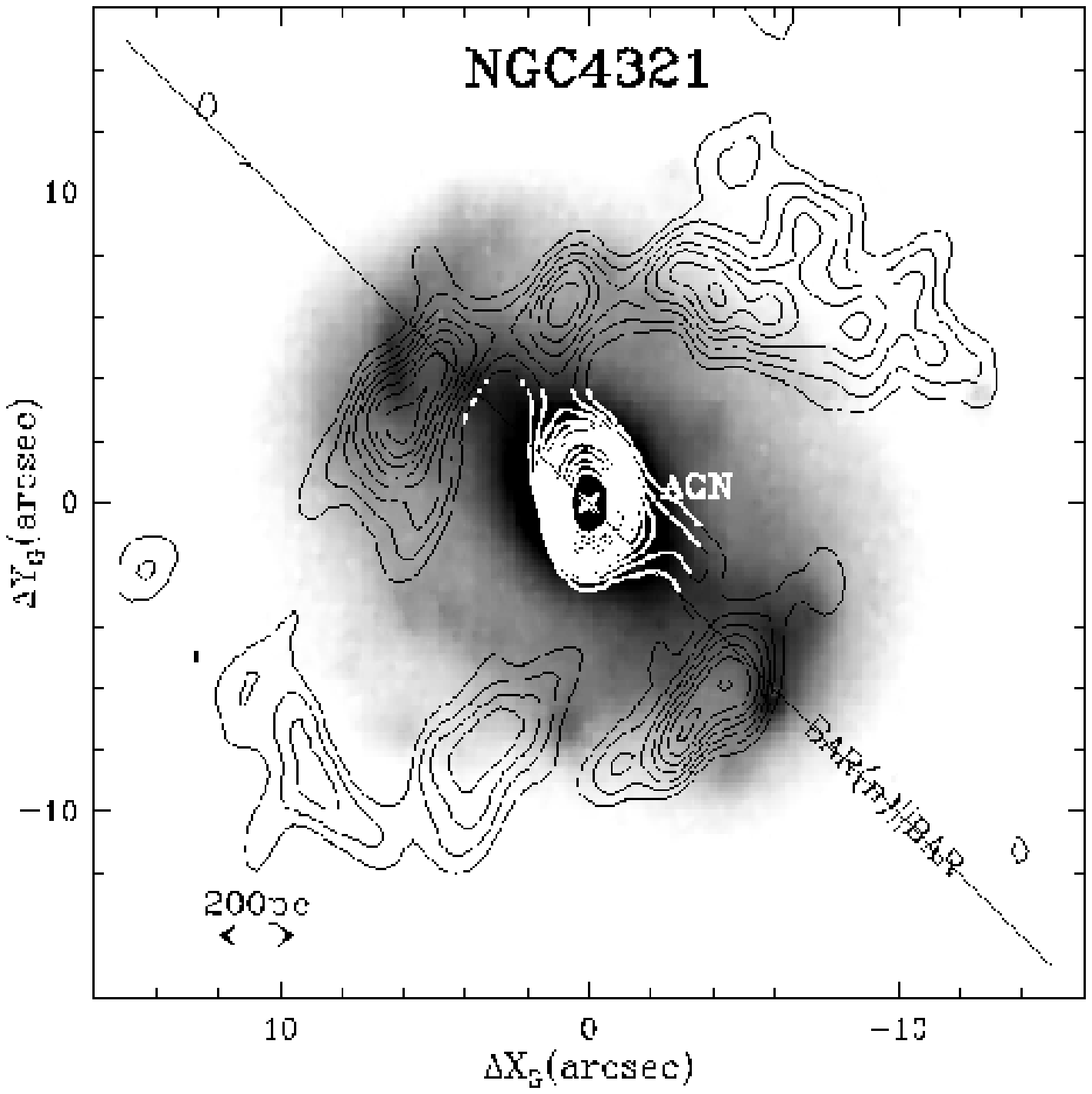}
   \includegraphics[width=8.2cm]{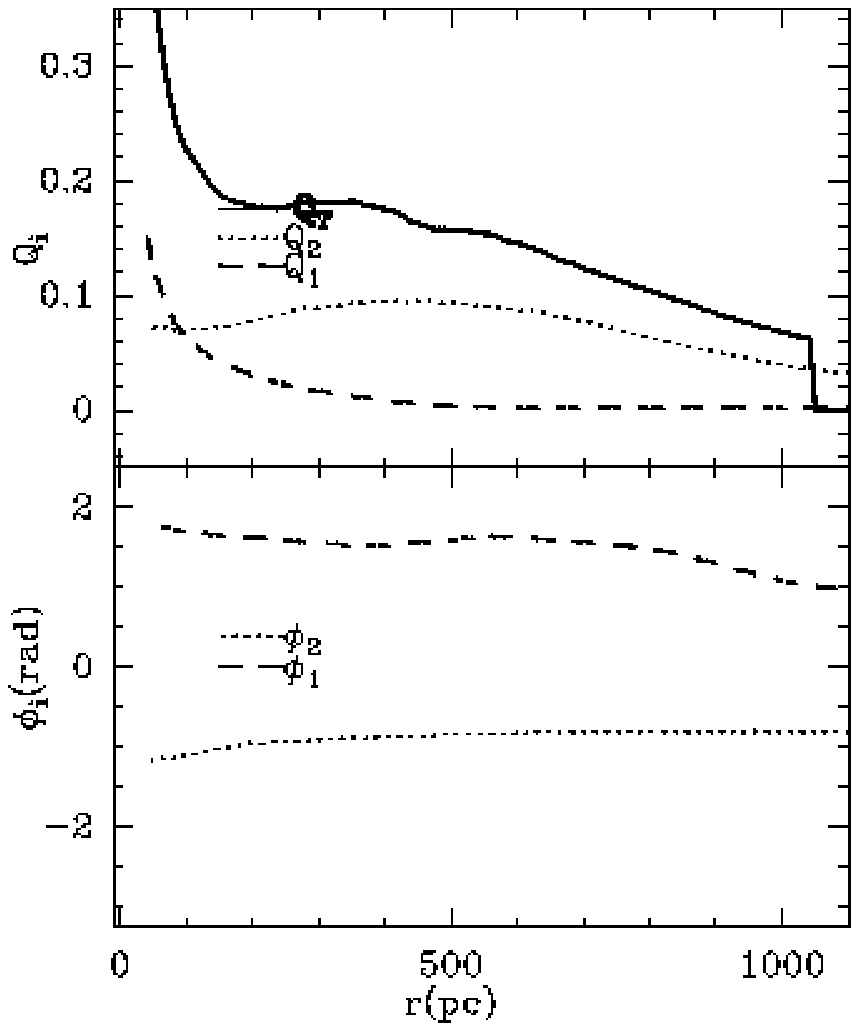}
   \caption{{\bf a) (upper panel)} We overlaid the $^{12}$CO(1--0) PdBI intensity map (contours) on the K-band 
image of Knapen et al.~(\cite{kna95}) (grey scale) obtained for the nucleus of NGC~4321; both images have been deprojected 
onto the galaxy plane. Units on X/Y axes ($\Delta$X$_G$/$\Delta$Y$_G$) correspond to arcsec offsets along the 
major/minor axes with respect to the AGN. The nuclear bar--BAR(n)--is parallel to the large-scale bar--BAR. {\bf b) 
(lower panels)} Strengths (Q$_i$, $i$=$1,2$) and phases ($\phi_i$, $i$=$1,2$) of the $m=1$ and $m=2$--Fourier 
components of the stellar potential inside the image field-of-view (r=1100~pc). The $\phi_i$-angles are measured from
the +X axis in the counter-clockwise direction. The total strength of the 
potential is represented by Q$_T$. Q-values for r$<$50~pc (not shown) are blanked due to insufficient grid 
sampling close to the nucleus. 
   }
         \label{fig:pot-4321}
 \end{figure}
 
Depending on the resolution/sensitivity requirements, we use either naturally or uniformly
weighted line maps, as indicated throughout the paper. Uniform weighting in the 2--1 line enables
us to achieve subarcsecond spatial resolution in the maps of
NGC\,4826, NGC\,4579 and NGC\,6951. By default, all velocities are referred to the systemic
velocities (v$_{sys}$, listed in Table~\ref{parameters}), as determined from this work and from Garc\'{\i}a-Burillo et
al.~(\cite{gb98}). Similarly ($\Delta\alpha$, $\Delta\delta$) offsets are relative to the AGN loci
derived from our own estimates (Table~\ref{parameters}).

Molecular gas masses are derived from the CO(1--0) integrated intensities assuming a CO--to--H$_2$ conversion factor X=N(H$_2$)/I$_{CO(1-0)}$=2.2$\times$10$^{20}$cm$^{-2}$~K$^{-1}$~km$^{-1}$~s (Solomon \& Barrett~\cite{sol91}). When required, molecular gas column densities are inferred from the CO(2--1) integrated intensity maps. In this case CO(2--1) intensities are first corrected by the 2--1/1--0 ratio measured within the equivalent 1--0 beam at each position, and subsequently we convert them into N(H$_2$) assuming the X factor referred to above.

We have estimated the percentage of CO(1--0) flux recovered in the PdBI maps by comparing the single-point fluxes 
detected by the 30m telescope (Garc\'{\i}a-Burillo \& Krips, private communication) towards the nuclei of the four galaxies discussed in this paper with the fluxes recovered in the PdBI maps, corrected by primary beam attenuation and convolved to the 30m resolution at this frequency (21$\arcsec$). The fraction of the flux recovered inside the 21$\arcsec$ field-of-view ranges from 60$\%$-65$\%$ in NGC~6951 and NGC~4579 to 75$\%$ 
in NGC~4321 and 90$\%$ in NGC~4826. The corresponding values for the 2--1 line of CO are comparable. The missing zero spacing
flux in these maps is expected to be found in low-level emission arising in the shape 
of smooth extended components.  As has been shown by Helfer et al.~(\cite{hel03}), who estimated the percentage of flux recovery in BIMA SONG galaxies using the 12m NRAO telescope, this percentage is always very high in the central regions of
galaxies (i.e., the domain of NUGA maps). The reason is that the velocity gradient is largest at the
nuclear regions. CO emission is thus confined to much smaller areas in individual channel maps compared to the outer
disk regions where the percentage of flux filtered out can be higher.
Considering that the percentage of flux actually present in the PdBI maps over the total single-dish estimate is moderate--to--large in the galaxies studied here (see above), we do not expect that the morphology of the maps will significantly change by the addition of a plateau-like component.  Moreover, the gravity torque calculation developed in this paper is based on azimuthal averages made to infer time-scales for the gas flows in these galaxy nuclei. The approach followed makes our results virtually insensitive to the presence of a weak extended component: gravity torques on this type of source distribution will be zero. Therefore we do not expect the derived torque budget to be significantly biased.

 \begin{figure}[tbh!]
   \centering
   \includegraphics[width=8.5cm]{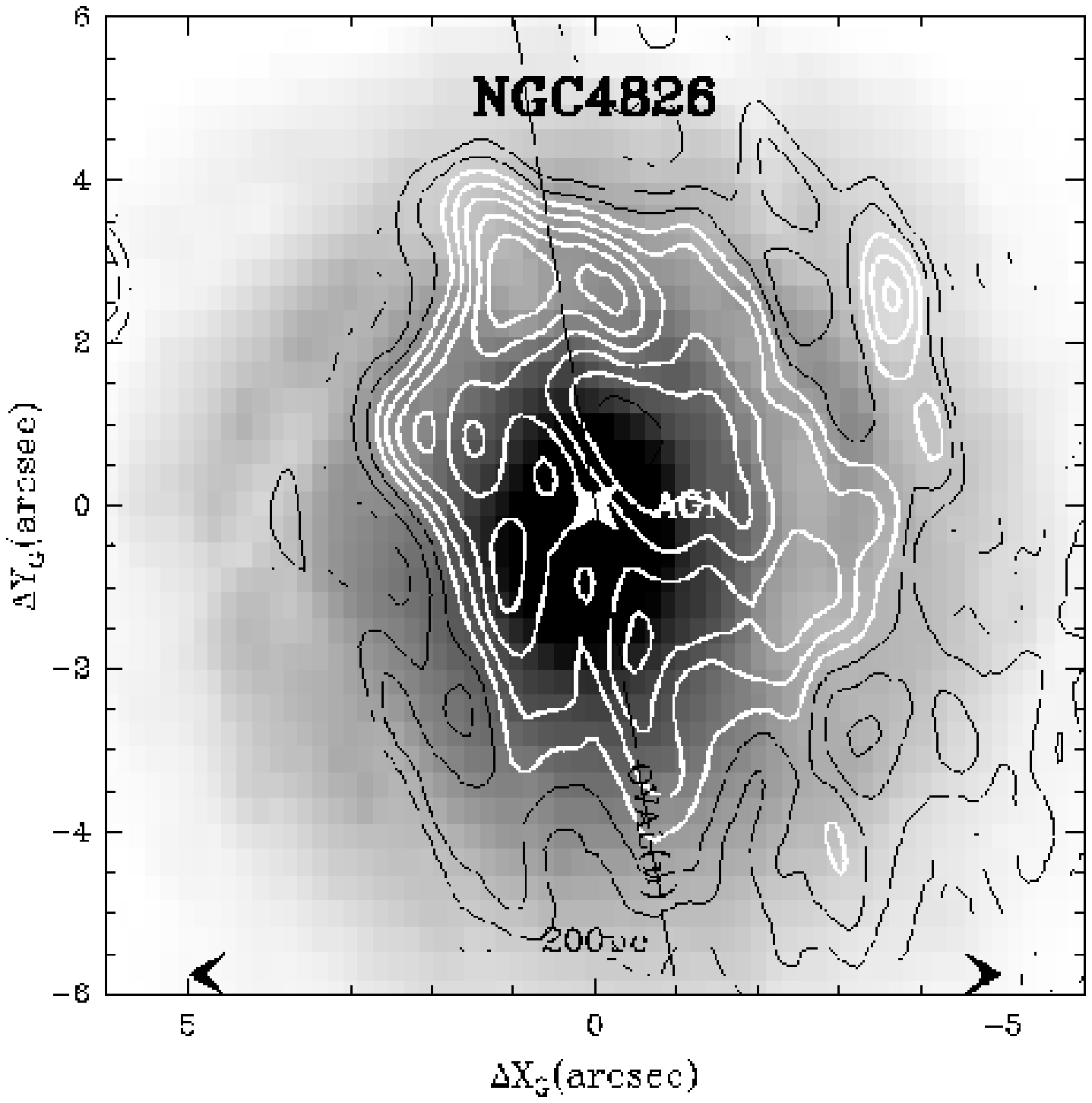}
   \includegraphics[width=8.2cm]{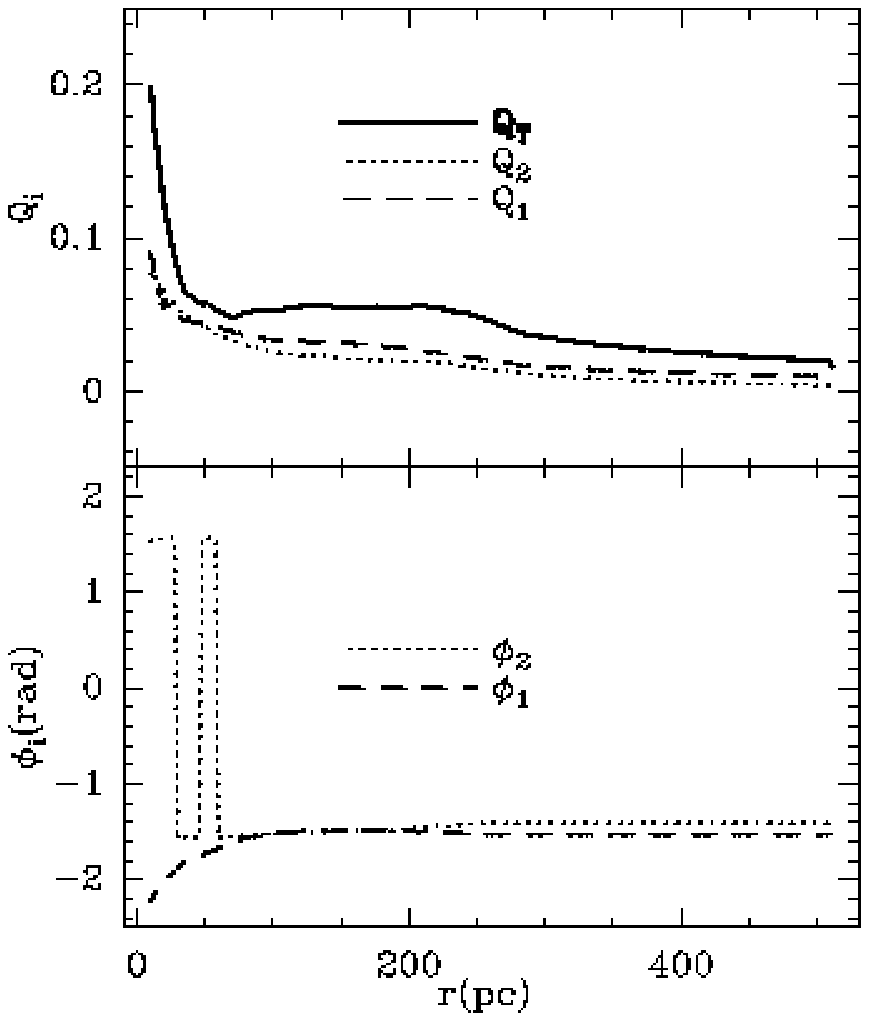}
   \caption{{\bf a) (upper panel)} We overlaid the $^{12}$CO(2--1) PdBI intensity map (contours) on the H-band HST 
image (grey scale) obtained for the nucleus (r$<$200~pc) of NGC~4826; both images have been deprojected onto the 
galaxy plane. The orientation of the nuclear oval perturbation--OVAL(n)--is shown. {\bf b) (lower panels)} As in 
Fig.~\ref{fig:pot-4321}, we plot (Q$_i$, $i$=$1,2$), ($\phi_i$, $i$=$1,2$) and  Q$_T$ inside the image 
field-of-view (r=530~pc) with an inner truncation radius r=10~pc. 
   }
         \label{fig:pot-4826}
 \end{figure}

\begin{figure}[tbh!]

   \centering
   \includegraphics[width=8.5cm]{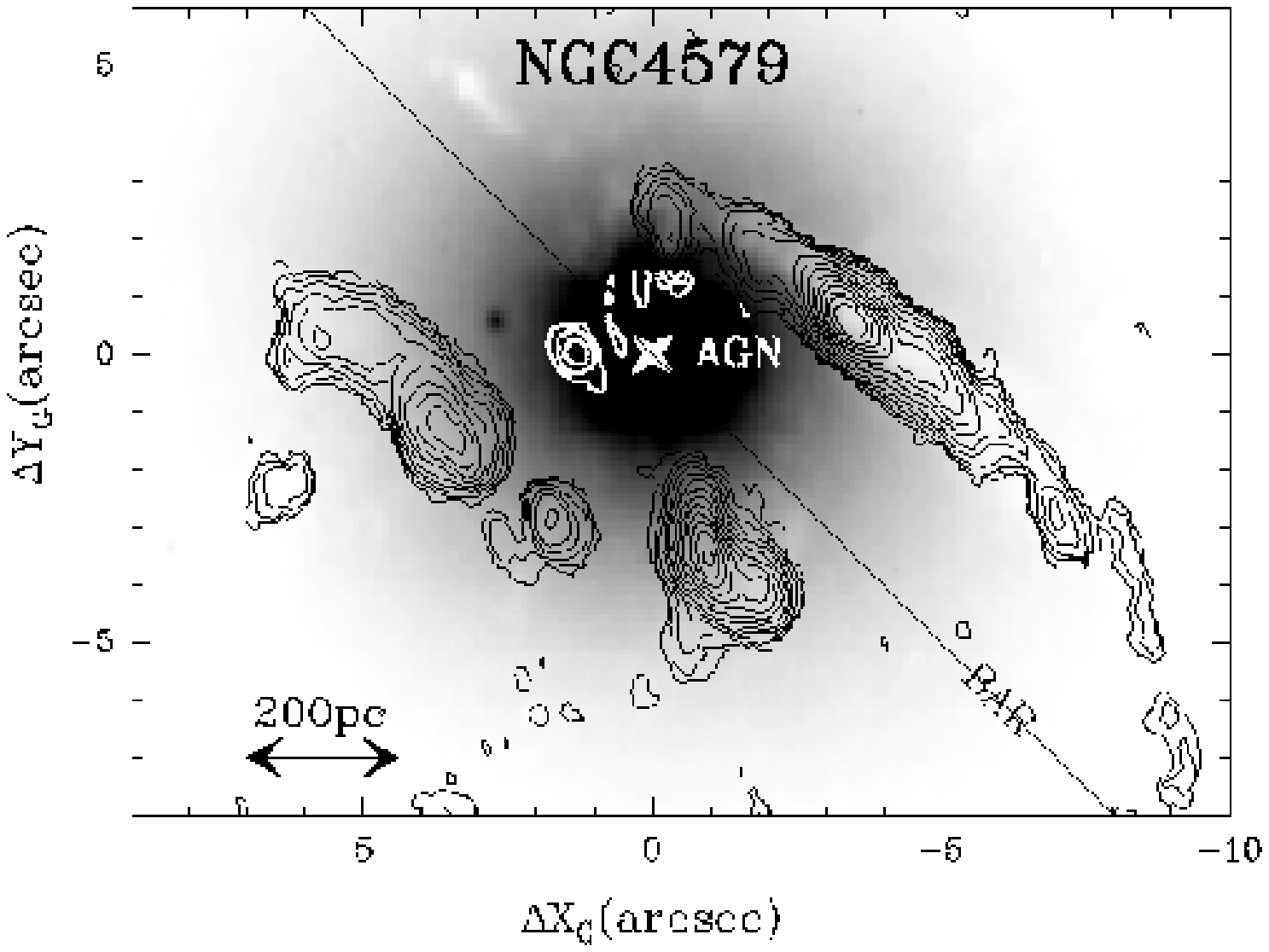}
   \includegraphics[width=8.2cm]{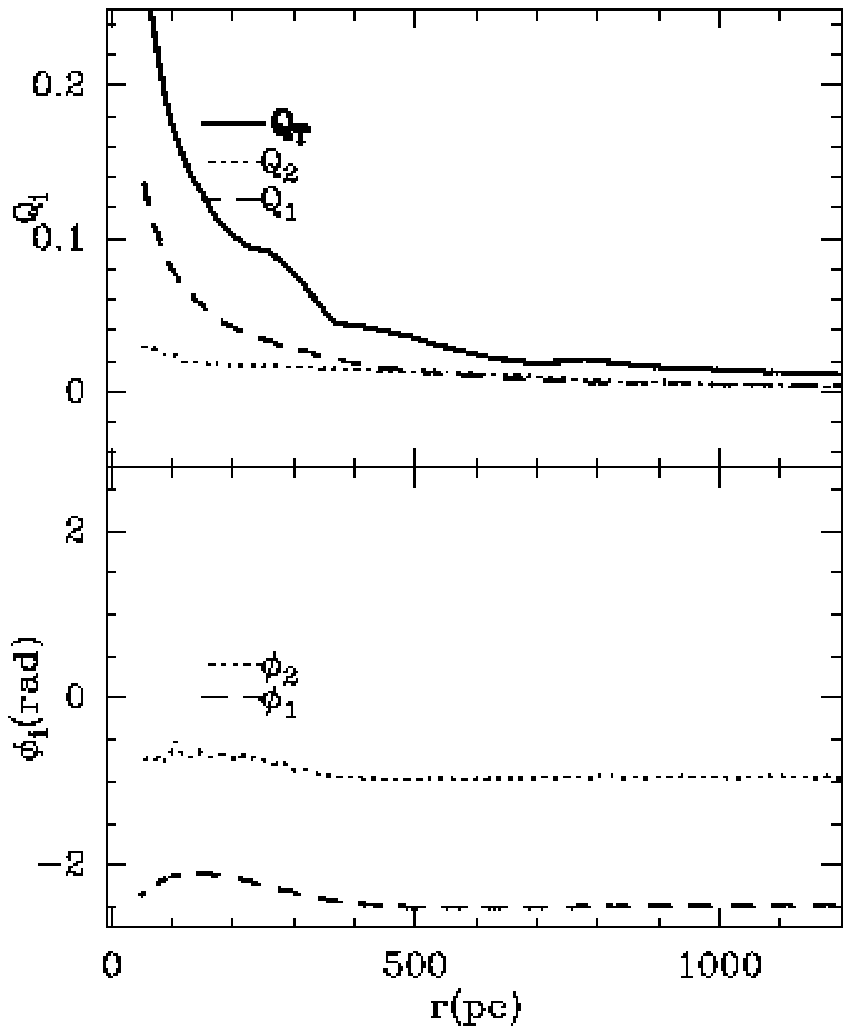}
   \caption{{\bf a) (upper panel)} We overlaid the $^{12}$CO(2--1) PdBI intensity map (contours) on the I-band HST 
image (grey scale) obtained for the nucleus of NGC~4579; both images have been deprojected onto the galaxy plane. 
The orientation of the large-scale 9~kpc bar--BAR--is shown.
   {\bf b) (lower panels)} As in Fig.~\ref{fig:pot-4321}, we plot (Q$_i$, $i$=$1,2$), ($\phi_i$, $i$=$1,2$) and  
Q$_T$ inside r=1200~pc with an inner truncation radius r=50~pc. 
   }
         \label{fig:pot-4579}
 \end{figure}
 
\begin{figure}[tbh!]
   \centering
   \includegraphics[width=8.5cm]{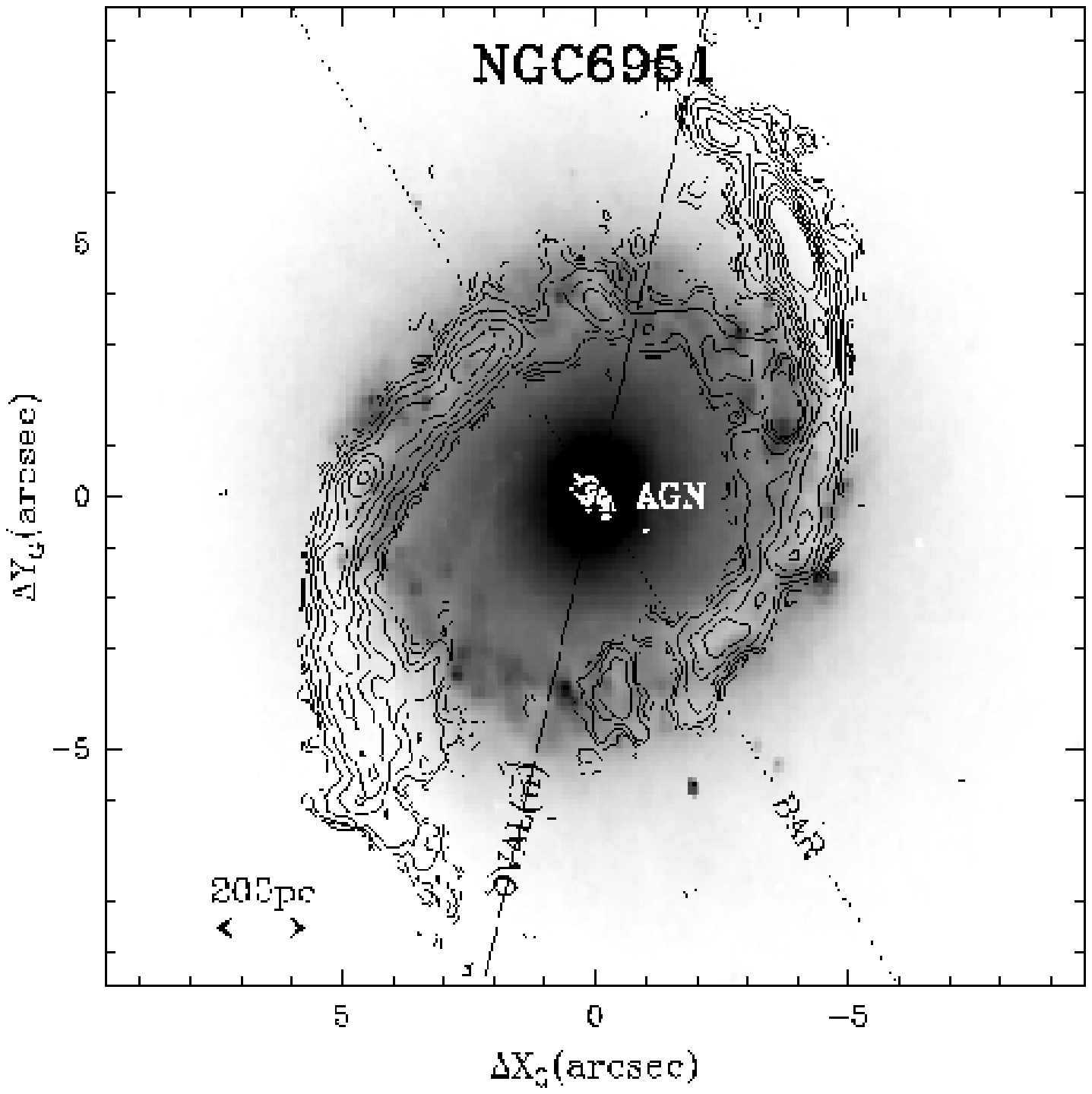}
   \includegraphics[width=8.2cm]{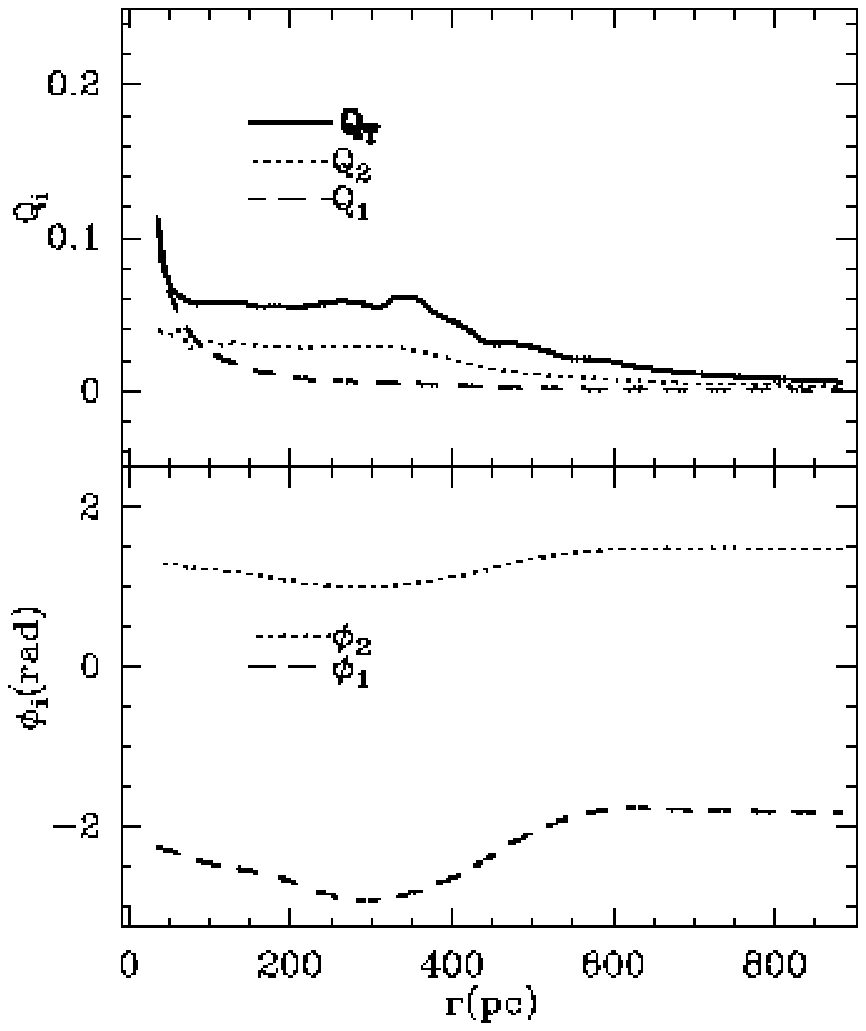}
   \caption{{\bf a) (upper panel)} We overlaid the $^{12}$CO(2--1) PdBI intensity map (contours) on the J-band HST 
image (grey scale) obtained for the nucleus of NGC~6951; both images have been deprojected onto the galaxy plane. 
We indicate the orientation of the large-scale bar--BAR--and of the nuclear oval distortion--OVAL(n)--that is 
detected in the HST image at r$<$500~pc. 
   {\bf b) (lower panels)} As in Fig.~\ref{fig:pot-4321}, we plot (Q$_i$, $i$=$1,2$), ($\phi_i$, $i$=$1,2$) and  
Q$_T$ inside the image field-of-view (r=900~pc) with an inner truncation radius r=25~pc. 
   }
         \label{fig:pot-6951}
 \end{figure}

\subsection{Near-infrared and optical observations \label{obshst}}

We acquired from the HST archive\footnote{Based on observations made with the NASA/ESA Hubble Space Telescope, 
obtained from the data archive at the Space Telescope Science Institute. 
STScI is operated by the Association of Universities for Research in 
Astronomy, Inc. under NASA contract NAS 5-26555.} broadband images of NGC\,4826, NGC\,4579 and NGC\,6951, including
three NICMOS images (F110W and F160W for NGC\,6951; F160W for NGC\,4826) and four WFPC2 images (F450W and F814W
for NGC\,4826; F555W and F814W for NGC\,4579). The optical images were combined using ({\it crreject}) to
eliminate cosmic rays, and calibrated according to Holtzman et al.~(\cite{holtzman}). The NICMOS
images were re-reduced with the STSDAS task {\it calnica} using the best reference files, and the
images were calibrated in the standard way. The ``pedestal'' effect (see B\"oker et
al.~\cite{bok99}) was removed with the van der Marel algorithm\footnote{\it
http://www.stsci.edu/$_{\tilde{\,}}$marel/software/pedestal.html}. Sky values were assumed to be
zero since the galaxy filled the WFPC2/NICMOS frames, an assumption which makes an error of
$\sim$\,0.1 mag at most, in the corner of the images. A J-H color image of NGC~6951 was constructed from
F110W-F160W according to the transformations by Origlia \& Leitherer~(\cite{origlia}). For
NGC\,4321, we have adopted the (ground-based) K-band image presented by Knapen et al.~(\cite{kna95}).


\section{Observational evidence of ongoing feeding\label{obs-feeding}}
 
\subsection{NGC\,4321\label{obs-n4321}}

The inner r$\sim$1.5~kpc of this galaxy was mapped by
Garc\'{\i}a-Burillo et al.~(\cite{gb98}) with the PdBI at  moderate ($\sim$2'') spatial resolution
in the 1--0 line emission of $^{12}$CO. NGC 4321 has 
been classified as a {\it transition} object (i.e., HII/LINER) by Ho et al.~(\cite{ho97}). As can be seen in 
Fig.~\ref{fig:feeding-4321}\textit{a}, molecular gas in the nucleus of NGC\,4321 is concentrated in a two 
spiral arm structure that starts at r$\sim$550~pc, near the end points of a prominent nuclear bar (detected in the K 
band by Knapen et al.~\cite{kna95}; see also Fig.~\ref{fig:pot-4321}) and extends out to r$\sim$1.2~kpc. There 
is also a central CO source coinciding with the AGN which is marginally resolved (r$\sim$150~pc) by the $\sim$2'' beam. 
This central source contains a molecular gas mass of 
$\sim$10$^{8}$M$_{\odot}$. The CO spiral arms mostly lie at the {\it trailing} edges of the nuclear bar. This 
particular geometry determines the feeding budget for the gas in this region (see Sect.~\ref{effi-n4321}). The 
spiral arms and the central source are connected by a molecular gas bridge which is spatially resolved NW of the 
nucleus, i.e., at the leading edge of the nuclear bar (component {\bf B} in Fig.~\ref{fig:feeding-4321}\textit{b}). 
Garc\'{\i}a-Burillo et al.~(\cite{gb98}) interpreted these CO observations and their 
relation with other gaseous and stellar tracers using numerical simulations of the cloud hydrodynamics. They found 
that the best fit for the gas flow corresponds to the nuclear bar being decoupled from the large-scale bar. The 
nuclear bar is {\it fast}, with a pattern speed $\Omega_p\geq$150~km~s$^{-1}$kpc$^{-1}$. This pushes corotation of 
the nuclear bar inward, likely inside the outer edges of the CO spiral arms. 

The kinematics of molecular gas are characterized by streaming motions detected in the CO spiral arms (see discussion 
in Garc\'{\i}a-Burillo et al.~\cite{gb98}). Closer to the AGN (r$<$500~pc), we find that molecular gas also 
displays significant departures from circular rotation at the location of {\bf B}.  Fig.~\ref{fig:feeding-4321}\textit{b} shows 
the position-velocity (p-v) plot along a strip at PA=127$^{\circ}$, purposely oriented to illustrate the CO 
kinematics at {\bf B}. 
The discontinuity in radial velocities at $\Delta$x$\sim$--4'' indicates that the molecular gas flow is decelerated, 
exactly as expected if gas flows along the leading edges of the bar.
Details on the kinematics of molecular gas in the central component are hidden due to the insufficient 
spatial resolution of the CO maps. We notice however that molecular gas emission is asymmetric with respect to 
v$_{sys}$: CO emission is preferentially blue-shifted. 

As is the case for the other {\it transition} object analyzed in this paper (NGC\,4826), we have found a large 
molecular gas concentration ($\sim$10$^{8}$M$_{\odot}$) near the AGN in NGC\,4321 (r$<$150\,pc). The analysis of 
gas kinematics provides evidence for gas fueling at present on intermediate scales: at r$\sim$200-300~pc from the 
AGN. However, the spatial resolution of the CO maps does not allow us to probe closer than 100~pc from the AGN.

\subsection{NGC\,4826\label{obs-n4826}}
The first NUGA maps of the {\it transition} object NGC\,4826 were published in paper I. The CO images showed 
already  a large concentration of molecular gas ($\sim$1600M$_{\sun}$/pc$^{2}$ ) in the $\sim$160~pc-diameter 
circumnuclear disk (CND) of this galaxy. The distribution of molecular gas in the inner CND is significantly 
lopsided with respect to the position of the AGN; this suggests that m=1 instabilities may be at work at radial 
distances of r$\sim$50--60~pc from the central engine. With the newest 0.5$^{\prime\prime}$ (10~pc) resolution 
$^{12}$CO(2--1) observations the distribution of molecular gas in the CND is fully resolved; the CND appears in 
the new maps as an off-center ringed disk (Fig.~\ref{fig:feeding-4826}\textit{a}). The dynamical center of the 
galaxy determined from these observations coincides within the errors with a blue point source identified in the 
B-I HST color map of Fig.~\ref{fig:feeding-4826}\textit{a}. This confirms our earlier findings that the putative 
super massive black hole lies on the southeastern inner side of the off-center ringed disk. 

The gas kinematics in the CND are characterized by the presence of streaming motions. A first analysis of the 2D 
kinematics performed on the lowest resolution data of paper I indicated that the instabilities identified in the 
CND (and those of the inner m=1 spiral) may not favor AGN feeding. The information contained in the newest images confirms this result.  Fig.~\ref{fig:feeding-4826}\textit{b} shows the departures from circular motion of gas velocities, identified at the northern and southern crossings of the CND ring along the minor axis. Gas velocities become systematically redder (bluer) when we approach the nucleus from the southern (northern) side of the ring along the minor axis. Deprojected onto the galaxy plane (North is the near side), this pattern indicates that the radial velocity component changes from an inflow signature (outside the ring) into an outflow signature (inside the ring). As it is fully discussed in paper I, this measured change of sign across the minor axis is compatible with the pattern expected for a trailing wave outside corotation ({\it fast} wave), i.e., the type of perturbation that would not help to drain the gas angular momentum.    
The driving agent of these 'mainly' gaseous $m=1$ instabilities may not be related to the stellar potential which is essentially featureless and mostly axisymmetric (see Sect.~\ref{effi-n4826} and discussion of paper I).

Based on the CO maps of NGC\,4826 we find little evidence of ongoing AGN
feeding at scales r$<$150\,pc from the AGN. While the molecular gas reservoir of NGC\,4826 is
abundant ($>$3$\times$10$^{7}$M$_{\odot}$) close to its central engine (r$<$150~pc), the analysis of
gas kinematics provides no evidence that ongoing AGN feeding is at work in this {\it transition}
object.

\subsection{NGC\,4579\label{obs-n4579}}

The $^{12}$CO(2--1) emission in the S1.9/L1.9 galaxy NGC\,4579 has been mapped at $\sim$0.5$^{\prime\prime}$ 
resolution (see Fig.~\ref{fig:feeding-4579}\textit{a} adapted from Garc\'{\i}a-Burillo et al.~2005 in prep.). The molecular gas distribution in the central $\sim$1~kpc of NGC\,4579 suggests that the gas flow responds to the 
9~kpc-diameter stellar bar identified in all the NIR images of this galaxy (e.g.~Jarret et al.~\cite{jar03}).  The $^{12}$CO(2--1) map of 
Fig.~\ref{fig:feeding-4579}\textit{a} reveals a mass of 3.2$\times$10$^{8}$M$_{\odot}$ of molecular gas piled up 
in two highly contrasted spiral arcs. The CO lanes lie at the leading edges of the stellar bar, which is oriented 
along PA=58$^{\circ}$. The northern spiral is more continuous and better delineated than its southern counterpart; 
it is also very well correlated with the red lane seen North in the V--I color HST image of the galaxy, 
shown in Fig.~\ref{fig:feeding-4579}\textit{a}. There is little molecular gas at r$<$100~pc distance from the central engine 
of NGC\,4579: the closest gas complex of $\sim$10$^{6}$M$_{\odot}$ lies East of the nucleus at r$\sim$150\,pc (complex E in Fig.~\ref{fig:feeding-4579}\textit{a,~b}). There is no molecular gas emission coincident with the AGN itself to a 
3$\sigma$ detection limit of $\sim$a few 10$^{5}$M$_{\odot}$. The V--I color HST image of the galaxy helps 
to identify the eastern  molecular complex as 
part of a structured disk of 150\,pc-diameter and ring-like shape (Fig.~\ref{fig:feeding-4579}\textit{a,~c}). The 
position of the AGN is well defined by its radio continuum emission detected at both 1mm and 3mm coming from a 
point source that lies close to the southwestern edge of the central disk. This indicates that the $m=2$ point-symmetry 
of the gas flow driven by the bar of NGC\,4579 breaks up at r$<$200\,pc and lets lopsidedness  take over. An 
independent confirmation of this picture comes from the new HST image of the nuclear region of NGC\,4579 
obtained with the ACS camera at 3300 \AA; this image resolves the central disk into a winding $m=1$ spiral 
instability that mimics a ring (Contini \cite{con04}).


The kinematics of molecular gas in the central 1~kpc of NGC\,4579 are characterized by the presence of highly
non-circular motions detected over the spiral arms and most notably over the central disk.
Fig.~\ref{fig:feeding-4579}\textit{b} shows the p-v plot along the kinematic major axis of NGC\,4579
(PA=95$^{\circ}$). The radial velocities of the gas in the central gas disk depart by $>$100kms$^{-1}$ from the
expected pattern of circular rotation: emission of the Eastern gas complex (at $\Delta$x$\sim$1.5$^{\prime\prime}$)
appears at {\it highly forbidden} negative velocities (i.e., v$<$v$_{sys}$). Assuming that the gas flows inside the
galaxy plane, the reported velocity deviations measured at E would imply that gas is {\it apparently}
counter-rotating at a speed of v$\sim$150kms$^{-1}$. As it is discussed in Garc\'{\i}a-Burillo et al.~2005, this could 
be qualitatively explained by very eccentric $m=1$ orbits. Alternatively, this velocity
pattern could be accounted for assuming that gas is flowing out of the galaxy plane, possibly entrained by an expanding
shell. The expanding shell scenario is supported by the observed kinematics of the gas close to the N side of the
central disk. Fig.~\ref{fig:feeding-4579}\textit{d} shows the p-v plot taken along the declination axis, i.e.,
very close to the orientation of the minor axis. Molecular gas kinematics at the N complex (where
v-v$_{sys}<$--75~kms$^{-1}$) can be interpreted either as gas flowing outward inside the plane along the minor axis
(North is the near side), or as a signature of out of the plane motions (similarly to the case of the E complex).
In either case this implies that AGN fueling is presently thwarted on these scales.

In summary, most of the molecular gas content of the central 1\,kpc of NGC\,4579 is trapped in a two
arm spiral structure that can be traced from r$\sim$1~kpc down to r$\sim$200~pc. Some molecular gas
(10$^{6}$M$_{\odot}$) is detected at r$\leq$150~pc from the central engine of NGC\,4579, but not on
the position of the central engine itself ($<$a few 10$^{5}$M$_{\odot}$). The first-order interpretation of the complex gas kinematics at r$<$150~pc provides no evidence of ongoing inflow from these scales down to the AGN, but on the contrary, it indicates
outflow motions.

\subsection{NGC\,6951\label{obs-n6951}}

NGC\,6951 is a prototypical Seyfert 2 galaxy for which sub-arcsecond resolution $^{12}$CO(2--1) maps have been 
completed within the NUGA project (Fig.~\ref{fig:feeding-6951}\textit{a} adapted from Schinnerer et al.~2005, in
prep.). The molecular gas distribution in the central 1~kpc consists of two nuclear spiral arms that can be traced
for over 180$^{\circ}$; the winding spiral arms end up as a highly contrasted  pseudo-ring at r$\sim$350~pc. The
spiral arms can be identified by their red color in the J--H HST image of the galaxy, shown in 
Fig.~\ref{fig:feeding-6951}\textit{a}. This $m=2$ gas instability  contains a significant gas reservoir of
3$\times$10$^{8}$M$_{\odot}$ which is presently feeding a nuclear starburst, also identified by its intense radio
continuum and H${\alpha}$ emissions (Ho \& Ulvestad \cite{ho01}, Rozas et al.~\cite{roz02}).  
The geometry of the molecular gas ridges likely reflects the crowding of molecular clouds along the x$_2$ family 
of orbits of the prominent stellar bar, detected in all NIR images of NGC\,6951 (M\'arquez \& Moles \cite{mar93}; 
Friedli et al.~\cite{fri96}; P\'erez et al.~\cite{per00}). As shown in 
Fig.~\ref{fig:feeding-6951}\textit{a}, only a small amount of molecular gas has succeeded in making its way down 
to the AGN: most of the molecular gas mass is trapped in the nuclear spiral arms, quite similar to the case of 
NGC\,4579 (though in this galaxy molecular gas likely populates both x$_1$ and x$_2$ orbits). A compact unresolved molecular 
complex (denoted as {\bf G} in Fig.~\ref{fig:feeding-6951}\textit{a, b}) of $\sim$a few 10$^{6}$M$_{\odot}$ is 
detected at the position of the central engine. This component could
correspond to a molecular torus (of $\sim$40-50~pc size). As seen in Fig.~\ref{fig:feeding-6951}\textit{a},
low level CO emission has been tentatively detected inside the ring, bridging the 'apparent' gap between the N
spiral arm (running North from West) and the central source. The bridge is better identified in the p--v plot of
Fig.~\ref{fig:feeding-6951}\textit{b}. The molecular gas mass of this emission bridge is more accurately estimated
using the natural weighted map: the gas mass amounts to $\sim$10$^{7}$M$_{\odot}$ (Schinnerer et al.~2005). This
could be the northern molecular counterpart of the filamentary spiral structure identified in the J--H HST map
(Fig.~\ref{fig:feeding-6951}\textit{a}).

The kinematics of molecular gas in the nuclear spiral arms reveal streaming motions (Schinnerer et al.~2005) 
also identified in the previous lower resolution CO and HCN maps of NGC\,6951 (Kohno et al.~\cite{koh99}).
Inside the ring (50~pc$<$r$<$350~pc), however, gas kinematics are compatible with regular rotation (Fig.~\ref{fig:feeding-6951}\textit{b}). Of particular note, this is in clear contrast to the case of NGC\,4579. The kinematics of the {\bf G}--component, although compatible with circular motions, cannot be studied in detail due to insufficient spatial resolution. 

While most of the 3$\times$10$^{8}$M$_{\odot}$ molecular gas disk is feeding a starburst episode in
the nuclear spiral arms at r$\sim$350~pc, a small amount of molecular gas ($\sim$a few 10$^{6}$M$_{\odot}$) has been
detected on the central engine. This central component reveals a prior accretion episode down to scales of r$\sim$50~pc.
It is unclear whether molecular gas detected in the bridge component is falling into the
nucleus or migrating outward: the gas flow is fully compatible with regular circular motions from r=50~pc to r=350~pc.

\section{Gravitational torques and AGN fueling\label{grav}}

To explore more precisely the efficiency of feeding, 
we have estimated the gravitational torques exerted by the stellar potentials (derived 
from the NIR images) on their molecular circumnuclear disks (as given by the NUGA CO maps). 
After estimating the role of stellar gravitational torques, 
we will investigate whether other mechanisms 
are required to explain the low level of nuclear activity in these galaxies.

  We first explain the general methodology employed and the basic assumptions in Sect.~\ref{method}. The different steps
are described in detail in Sect.~\ref{grav-calc} and  Sect.~\ref{torq-calc}. We discuss in Sect.~\ref{track} the results obtained from the application of this procedure to the four NUGA targets examined in this work.

\subsection{General methodology\label{method}}

Gravitational forces are computed at each location in the plane of the galaxy, using near-infrared images to derive the underlying gravitational potential. We assume that the total mass budget is dominated by the stellar contribution and thus neglect the effect of gas self-gravity. We also assume a constant M/L ratio, and determine its best value by fitting the rotation curve constrained by the CO observations.
From the 2D force field ($Fx$,$Fy$) we derive the torques per unit mass at each location ($t(x,y)$=$x~F_y-y~F_x$). This torque field, by definition, is independent of the present gas distribution in the plane. The crucial step consists of using the torque field to derive the angular momentum variations and the associated flow time-scales. As explained below, the link is made through the observed distribution of the gas. 

With this aim, we assume that the measured gas column density ($N(x,y)$) derived from a CO intensity map at each offset in the galaxy plane is a fair estimate of the probability of finding gas at this location at present. In this statistical approach, we implicitly average over all possible orbits of gaseous particles and take into account the time spent by the gas clouds along the orbit paths.
We assume that CO is a good tracer of the total gas column density, since HI mass in the nuclei of galaxies is typically
a very small fraction of the total gas mass. The torque field is then weighted by $N(x,y)$ at each location to derive the time derivative of the local angular momentum surface density $dL_s(x,y)/dt$=$N(x,y) \times t(x,y)$.

In order to estimate the gas flows induced by these angular momentum variations we produce azimuthal averages of $dL_s(x,y)/dt$ at each radius. The azimuthal average at each radius, using $N(x,y)$ as the actual weighting function, represents the global variation of the specific gas angular momentum occurring at this radius ($dL/dt~\vert_\theta$). Finally, the time-scales for gas inflow/outflow can be derived by estimating the average fraction of angular momentum transferred in one rotation.

The validity of our estimate of the efficiency of stellar gravity torques to drive angular momentum transfer in the gas is based on the following simple hypothesis: we assume that the gas response to the stellar potential is roughly stationary  with respect to the potential reference frame during a few rotation periods. We would like to stress that even in the particular case of nuclear bars, which might decouple from the outer stellar bars under certain circumstances, our assumption is still valid. When there are several stellar pattern speeds at different radii in a galaxy disk, numerical simulations show that the gas response tends to be coupled with the stellar potential pattern (Friedli \& Benz~\cite{fri95}; Garc\'{\i}a-Burillo et al.~\cite{gb98}; Bournaud \& Combes~\cite{bou02}, 2005 in prep.). The gas response adjusts its pattern speed to that of the dominant stellar pattern at a given radius.

A different case is represented by non-axisymmetric  perturbations which can 
be driven by gas self-gravity and that are partly independent of or possibly decoupled
from the stellar perturbations of the disk. Nuclear galaxy disks with a high gas surface 
density and a mostly axisymmetric stellar potential can be prone to develop
this kind of gas self-gravitating perturbation. In this limiting case (not 
contemplated here), and although our calculation is still formally correct, the inclusion of gas 
self-gravity is required to derive the correct torque budget. In particular, 
if the gas disk decouples from the stellar pattern, the azimuthally averaged 
gravity torques exerted by the stellar pattern on the gas will very likely be 
close to zero. The main source for the torques if any should 
come from the gas instability itself.

Note also that any radial variation of the CO-to-H$_2$ conversion factor (X$_{CO}$) is not expected to affect the estimated time-scales as these are independent of the global normalization factor as a function of radius. If X$_{CO}$ varies as a function of azimuth at a fixed radius, there could be a potential bias. However, observational evidence indicates that the 
dominant variation of X$_{CO}$ in the central regions of galaxies is radial (Solomon \& Barrett~\cite{sol91}; Regan et al.~\cite{reg01}).

In the approach followed to derive the gravitational potential we have assumed that the bulge is as flattened as the disk, and thus no attempt has been made to separate the bulge from the pure disk contribution. The implicit assumption of a
highly flattened bulge is probably not wrong for some barred galaxies but for others it will overestimate  
the radial forces by at most a factor of 2 (e.g., Buta \& Block~\cite{but01}). On the other hand, 
the effect of bulge stretching due to deprojection can enhance the strength of bars, especially if they are aligned with 
the minor axis and the inclination angles are large. As this is not the case for the galaxies analyzed here, we instead expect 
that the value derived for the gravity torques will be typically underestimated in our case by a factor 1.5-2 although they will still have the same sign.

\subsubsection{Evaluation of stellar potentials\label{grav-calc}}

The first step is to derive the stellar potential in the nuclear disks of these 
galaxies, using the high-resolution NIR images described in Sect.~\ref{obshst}. Our working
hypothesis is that NIR images are less affected than optical ones by 
dust extinction or by stellar population biases (Quillen et al.~\cite{quillen}). 
The images are first deprojected according to the angles $PA$ and $i$ given in Table~\ref{parameters}. 
The images are then completed in the vertical
dimension by assuming an isothermal plane model with a constant scale height,  
equal to $\sim$1/12th of the radial scale-length of the image. The potential is
then derived by a Fourier transform method. We also assumed a constant mass-to-light (M/L) ratio, 
obtained by fitting the observed CO rotation curve--$v_{rot}$--for each galaxy. The  
 potential--$\Phi(R,\theta)$-- is then decomposed in the different m-modes:

\begin{equation}
\Phi(R,\theta) = \Phi_0(R) + \sum_m \Phi_m(R) \cos (m \theta - \phi_m(R))
\end{equation}

\noindent
where $\Phi_m(R)$ and $\phi_m(R)$ represent the amplitude and phase of the m-mode, respectively.

Following Combes and Sanders~(\cite{com81}), we define the strength of the $m$-Fourier component,
$Q_m(R)$ as
 \begin{equation}
Q_m(R)=m \Phi_m / R | F_0(R) |
 \end{equation}
The corresponding strength of the total non-axisymmetric perturbation is defined by:
\begin{equation}
Q_T(R) = {F_T^{max}(R) \over F_0(R)} =
{{{1\over R}\bigl{(}{\partial \Phi(R,\theta)\over \partial\theta}\bigr{)}_
{max}} \over {d\Phi_0(R)\over dR}}
\end{equation}
\noindent
where $F_T^{max}(R)$ represents the maximum amplitude of the tangential force over all $\theta$ and $F_{0}(R)$ is the mean 
axisymmetric radial force.

Figs.~\ref{fig:pot-4321}\textit{a,b} to \ref{fig:pot-6951}\textit{a,b} illustrate the quantitative description of 
gravitational potentials given by [Q$_{i=1,2}$,~Q$_T$,~$\phi_{i=1,2}$] for the four galaxies examined in this 
work.

\subsubsection{Efficiency of gravitational torques\label{torq-calc}}

After having calculated the forces per unit mass ($F_x$ and $F_y$) from the derivatives of $\Phi(R,\theta)$ at 
each pixel, the torques per unit mass--$t(x,y)$--can be  computed by:

\begin{equation}
t(x,y) = x~F_y -y~F_x
\end{equation}

\begin{figure*}[tbh!]
   \centering
   \includegraphics[width=18cm]{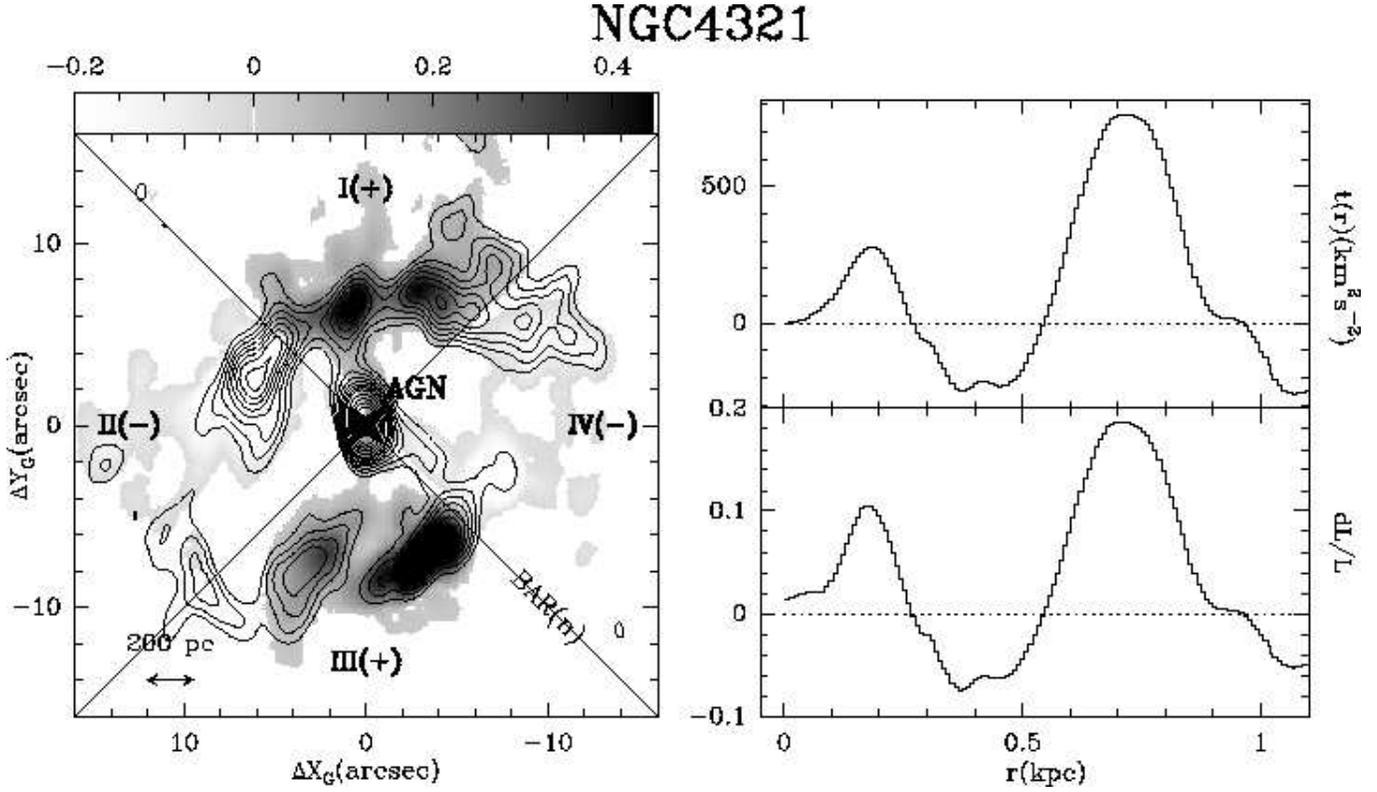}
   \caption{{\bf a)(left)}~We overlay the $^{12}$CO(1--0) contours with the map of the effective angular momentum variation (t(x,y)~$\times$~N(x,y), as defined in the text) in the nucleus of NGC~4321. The grey
scale is normalized to the maximum absolute value in the map, i.e.,
[$\vert$~N(x,y)~$\times$~t(x,y)$~\vert$]$_{max}$. The derived torques change sign as expected if the {\it butterfly} diagram, defined by the
orientation of quadrants I-to-IV, can be attributed to the action of the nuclear bar of NGC\,4321. {\bf
b)(right)}~The torque per 
unit mass averaged over azimuth--$t(r)$--and the fraction of the angular momentum transferred
from/to the gas 
in one rotation--$dL/L$--are plotted. Torques are strong and positive for the bulk of the molecular
gas in 
NGC\,4321. This includes the vicinity of the AGN. Torques are negative but comparatively weaker on
intermediate 
scales (r=250-550~pc) and in the outer disk r$>$900~pc.    
   }
         \label{fig:torques-4321}
 \end{figure*}

The sense of the circulation of the gas in the galaxy plane determines the sign of $t(x,y)$: positive (negative) if the torque 
accelerates (decelerates) the gas at $(x,y)$. We have next obtained the gravitational torque
maps weighted by the gas column densities derived from the CO 1--0 and 2--1 lines, $N(x,y)$, i.e., we derive $t(x,y)\times
N(x,y)$. These represent the effective variations of angular momentum density in the galaxy plane. We show in Fig.~\ref{fig:torques-4321}\textit{a} to \ref{fig:torques-6951}\textit{a} the normalized version of these maps, i.e., divided by [$\vert$~N(x,y)~$\times$~t(x,y)$~\vert$]$_{max}$ for the 2--1 line (except for NGC~4321, where we used the 1--0 line). To estimate the radial gas flow induced by the torques, we have first computed the torque 
per unit mass averaged over the azimuth, using $N(x,y)$ as the actual weighting function,i.e.:

\begin{equation}
t(R) = \frac{\int_\theta N(x,y)\times(x~F_y -y~F_x)}{\int_\theta N(x,y)}
\end{equation}

Results obtained for t(R) based on the CO(1--0) and CO(2--1) maps were seen to be virtually identical within the errors. Fig.~\ref{fig:torques-4321}\textit{b} to \ref{fig:torques-6951}\textit{b} show the results derived from the 1--0 line maps.
By definition, $t(R)$ represents the time derivative of the specific angular momentum--$L$--of the gas averaged 
azimuthally, i.e., $t(R)$=$dL/dt~\vert_\theta$. To derive azimuthal averages, we assume a radial binning ($\Delta R$) which 
corresponds to the original resolution of the NIR images.  Similarly to the
torque maps, the sign of 
$t(R)$, either + or --, defines whether the gas may gain or lose angular momentum, respectively. 
 More precisely, we evaluate 
the AGN feeding efficiency by deriving the average fraction of the gas specific angular momentum transferred in one rotation 
(T$_{rot}$) by the stellar potential, as a function of radius, i.e., by the non-dimensional function $\Delta L/L$ defined as:

\begin{equation}
{\Delta L\over L}=\left.{dL\over dt}~\right\vert_\theta\times \left.{1\over L}~\right\vert_\theta\times 
T_{rot}={t(R)\over L_\theta}\times T_{rot}
\label{tgrav}
\end{equation}

\noindent
where $L_\theta$ is assumed to be well represented by its axisymmetric average, i.e.,$L_\theta=R\times v_{rot}$. 
The absolute value of $L/\Delta L$ determines how long will it take for the stellar potential to transfer the 
equivalent of the total gas angular momentum. Assuming that the gas response to the stellar potential is 
stationary with respect to the potential reference frame during a few rotation periods, a small
value of $\Delta L / L$ implies that the stellar potential is inefficient at present. The $t(R)$ and
$\Delta L/L$ curves derived from the 1--0 maps of the analyzed galaxies are displayed in
Figs.~\ref{fig:torques-4321}\textit{b} to \ref{fig:torques-6951}\textit{b}.

\begin{figure}[tb!]
   \centering
   \includegraphics[width=8.5cm]{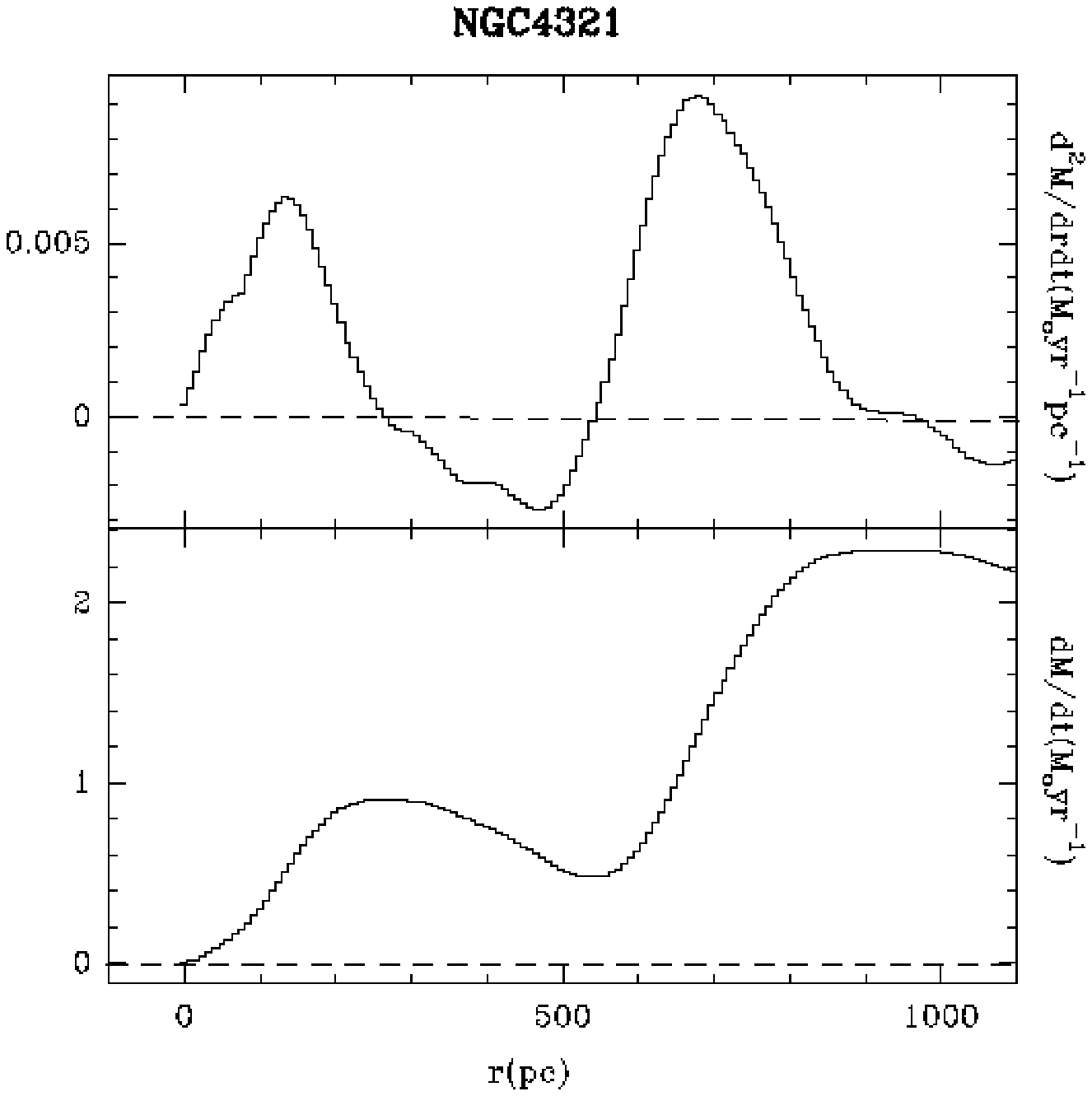}   
    \caption{{\bf a)(upper panel)} 
   We represent the radial variation of the mass inflow(-) or outflow(+) rate of gas per unit radial length in the 
nucleus of NGC\,4321 due to the action of stellar gravitational torques. Units are M$_{\sun}$~yr$^{-1}$pc$^{-1}$.
   {\bf b)(lower panel)} Here we plot the mass inflow/outflow rate integrated inside a certain radius r in 
M$_{\sun}$~yr$^{-1}$. 
   As is the case of the other transition object, NGC~4826, the overall budget in NGC\,4321 is clearly positive at 
all radii.}
         \label{fig:dm-4321}
 \end{figure}

To calculate how much gas mass is involved in the transfer driven by the stellar potential    
we have estimated the radial trend for the mass inflow (- sign)/outflow(+ sign) rate of gas per unit length as a 
function of radius (in units of M$_{\odot}$~yr$^{-1}$pc$^{-1}$ in 
Figs.~\ref{fig:dm-4321}\textit{a} to \ref{fig:dm-6951}\textit{a}) as follows: 

\begin{equation}
{d^2M\over dRdt}=\left.{dL\over dt}~\right\vert_\theta\times \left.{1\over L}~\right\vert_\theta\times 2\pi 
R\times \left.N(x,y)~\right\vert_\theta
\end{equation}
\noindent
where $\left.N(x,y)~\right\vert_\theta$ is the radial profile of N(x,y) averaged over the azimuth for a radial 
binning $\Delta R$. 
 
The inflow/outflow rates integrated out to a certain radius R can be derived as:
\begin{equation}
{dM\over dt}=\sum {d^2M\over dRdt}~\times\Delta R
\end{equation}

Figs.~\ref{fig:dm-4321}\textit{b} to \ref{fig:dm-6951}\textit{b} display these integrated rates in units of 
M$_{\odot}$~yr$^{-1}$.

\begin{figure*}[tbh!]
   \centering
   \includegraphics[width=18cm]{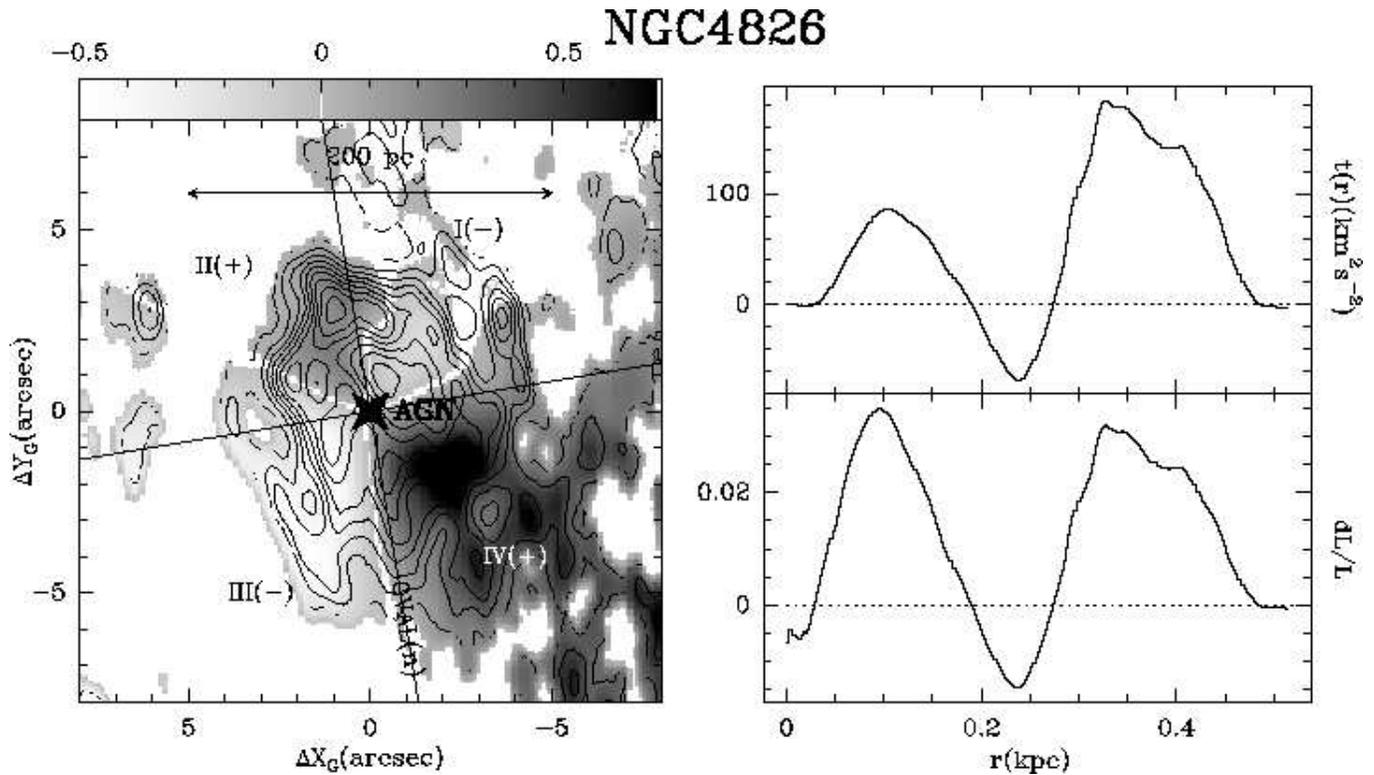}
    \caption{Same as Fig.~\ref{fig:torques-4321} but for NGC~4826, using the $^{12}$CO(2--1) map
in the overlay and the $^{12}$CO(1--0) map in the radial averages. The derived torques change 
sign as expected if the {\it butterfly} diagram, defined by the orientation of quadrants I-to-IV,
can be attributed to the action of the oval distortion of NGC\,4826. Stellar torques are weak and
mostly positive in the nucleus of NGC\,4826 at present.}
         \label{fig:torques-4826}
 \end{figure*}

 \begin{figure*}[tbhp!]
   \centering
   \includegraphics[width=18cm]{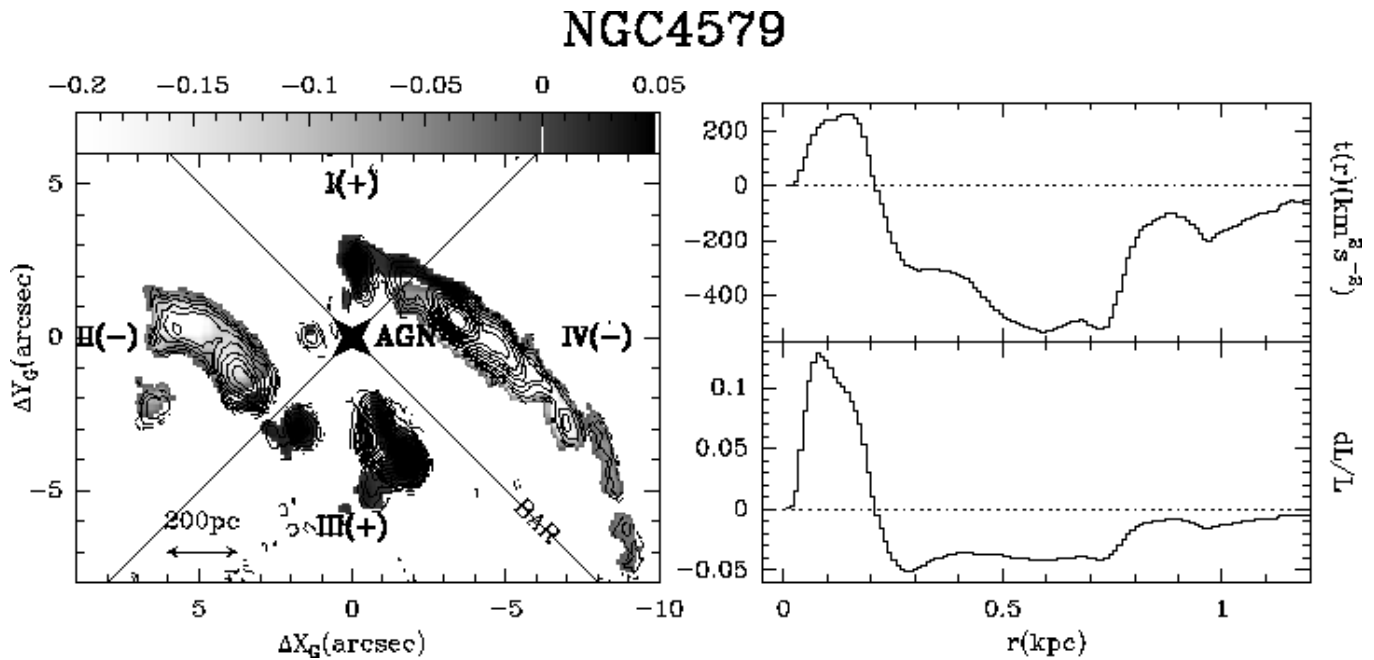}
   \caption{Same as Fig.~\ref{fig:torques-4826} but  for NGC~4579. The torques change sign as 
expected if the {\it butterfly} diagram, defined by the orientation of 
   quadrants I-to-IV, can be attributed to the action of the large-scale bar of NGC\,4579. Torques are 
systematically strong and negative for the bulk of the molecular gas in NGC\,4579, from r=200~pc out to r=1200~pc. 
In the vicinity of the AGN, however, torques become positive and AGN feeding is not presently favored.
              }
         \label{fig:torques-4579}
 \end{figure*}
 
\begin{figure*}[tbhp!]
   \centering
   \includegraphics[width=18cm]{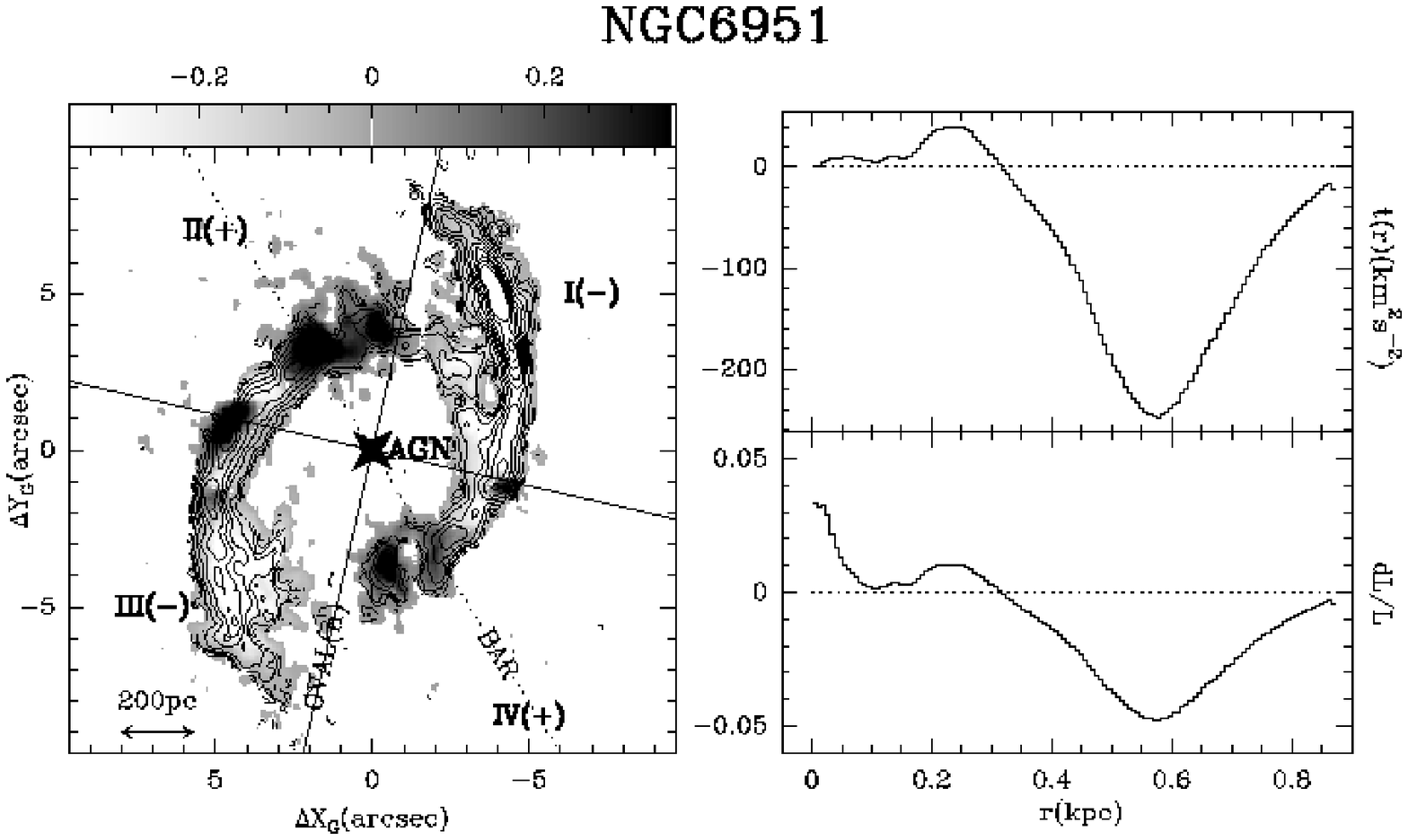}
   \caption{Same as Fig.~\ref{fig:torques-4826} but for NGC~6951. The torques change sign as 
expected if the {\it butterfly} diagram, defined by the orientation of 
   quadrants I-to-IV, can be attributed to the action of the nuclear oval rather than to that of the outer bar of 
NGC\,6951. Torques are systematically negative over the nuclear spiral arms down to the inner radius of the 
pseudo-ring, r$\sim$300~pc. In the vicinity of the AGN, torques become positive and AGN feeding is not 
presently favored.
              }
         \label{fig:torques-6951}
 \end{figure*}

\subsection{Tracking down gravitational torques in NUGA targets\label{track}}

\subsubsection{NGC\,4321\label{effi-n4321}}

Fig.~\ref{fig:pot-4321}\textit{a} shows the K-band image of the nucleus of NGC~4321, obtained by Knapen et 
al.~(\cite{kna95}), deprojected onto the galaxy plane (X$_G$/Y$_G$ coordinates). The NIR image shows a nuclear bar 
(denoted as BAR(n) in Fig.~\ref{fig:pot-4321}\textit{a}) which is roughly parallel to the large-scale 8~kpc 
stellar bar (oriented along PA=112$^{\circ}$ in sky coordinates). The nuclear bar strength is
the main contributor to Q$_{T}$, which is typically $\sim$0.18 for r$<$500~pc (Fig.~\ref{fig:pot-4321}\textit{b}).

Fig.~\ref{fig:torques-4321}\textit{a}, shows that the derived torques change sign following a characteristic 2D 
{\it butterfly} pattern. 
Quadrants I-to-IV will define hereafter the regions where the signs of the torques driven by the
dominant perturbation of the stellar potential are expected to be constant 
(Figs.\ref{fig:torques-4321}-\ref{fig:torques-6951}\textit{a}). 
Assuming that the gas circulation is counterclockwise, 
Fig.~\ref{fig:torques-4321}\textit{a} shows that the bulk of the CO emission along the spiral arms lies at the 
trailing edges of the nuclear bar where torques are positive (quadrants I(+) and III(+)). While there is 
significant CO emission arising from the leading quadrants of the bar (quadrants II(-) and IV(-)) both on 
intermediate scales (r=250-550~pc) and in the outer disk r$>$900~pc, negative torques there are weaker than 
positive torques measured over the trailing quadrants (see Fig.~\ref{fig:torques-4321}\textit{b}). Most 
remarkably, closer to the AGN (r$\leq$200~pc), the sign of the average torques $t(R)$ is also positive. Although 
the central CO source is marginally resolved in the CO map of Garc\'{\i}a-Burillo et al.~(\cite{gb98}), and, 
therefore, higher resolution observations are required to settle the question, positive torques seem to dominate 
over negative ones at r$\leq$200~pc.

In terms of mass inflow rates, the overall feeding budget in NGC\,4321 is clearly {\it positive} at all radii as 
shown in Fig.~\ref{fig:dm-4321}. This would imply that the radial gas flow goes outward unless other mechanisms 
 more efficiently transfer the gas angular momentum outwards. 
At r$\sim$100~pc, dM/dt$\sim$+0.4M$_{\sun}$~yr$^{-1}$ and the radial rate is still {\it positive} at
r$\sim$500~pc, where dM/dt$\sim$+0.4M$_{\sun}$~yr$^{-1}$. Similarly to the case of the other transition object, 
NGC~4826 (see Sect.~\ref{effi-n4826}), AGN feeding driven by stellar torques is quenched at present in NGC\,4321. Although the gas 
concentration is large close to the AGN, the present gas configuration favors feeding only on intermediate scales 
(r=250-550~pc). This result is compatible with the CO-based diagnostic of the gas kinematics discussed in 
Sect.\ref{obs-n4321}.   

\subsubsection{NGC\,4826\label{effi-n4826}}

Fig.~\ref{fig:pot-4826}\textit{a} shows a zoom into the central $\sim$200~pc region of the deprojected H-band HST 
image of NGC~4826. We plot in Fig.~\ref{fig:pot-4826}\textit{b} Q$_i$( $i$=$1,2$), $\phi_i$ ($i$=$1,2$) and  Q$_T$ 
inside the full field-of-view of the image (r=530~pc). As shown in Fig.~\ref{fig:pot-4826}\textit{b}, the 
deviations from axisymmetry of the stellar potential are exceedingly small at all radii (Q$_T\sim$0.02--0.05) 
except for r$<$75~pc where a weak oval perturbation is detected. The latter is denoted as OVAL(n) in 
Fig.~\ref{fig:pot-4826}\textit{a}. This oval seems slightly off-center with respect to the AGN; this would 
account for the contribution from a $m=1$ stellar mode to Q$_{T}$, identified in Fig.~\ref{fig:pot-4826}\textit{b}.

As already suggested by the first calculations of paper I, we find that the weak stellar perturbations present in 
the inner 500~pc of NGC~4826 are inefficient at feeding the AGN at present. Fig.~\ref{fig:torques-4826}\textit{a} 
shows how the derived torques change sign in the central 200~pc of the disk. The distribution of torques, defined 
by quadrants I-to-IV in Fig.~\ref{fig:torques-4826}\textit{a}, shows the expected behavior if torques are mostly 
due to the oval distortion of NGC\,4826. Moreover, stellar torques are weak and mostly positive inside 
r$\sim$150~pc (Fig.~\ref{fig:torques-4826}\textit{b}). Very close to the AGN (r$\leq$20~pc), the sign 
of average torques might become marginally negative, but as we lack spatial resolution on these
scales, the torque estimate becomes uncertain. 

Alternatively, the lopsided gas disk may be unrelated to the stellar oval. 
Instead, gas self-gravity could be driving the gas response in the nucleus NGC~4826 
(see discussion in paper I). In this case the gas disk may be decoupled from the 
stellar pattern, and therefore the latter would be viewed as a mostly axisymmetric 
component from the system of reference of the gas, i.e., providing zero torque on 
the gas disk. The main source for the torques should come from the gas 
instability itself. In NGC~4826 $m=1$ instabilities do not seem to favour gas inflow, however (paper I).
In either case, this is compatible with the scenario depicted in  Sect.~\ref{obs-n4826}, 
in which there is currently no evidence of fueling of the AGN. 

\begin{figure}[tb!]
   \centering
   \includegraphics[width=8.5cm]{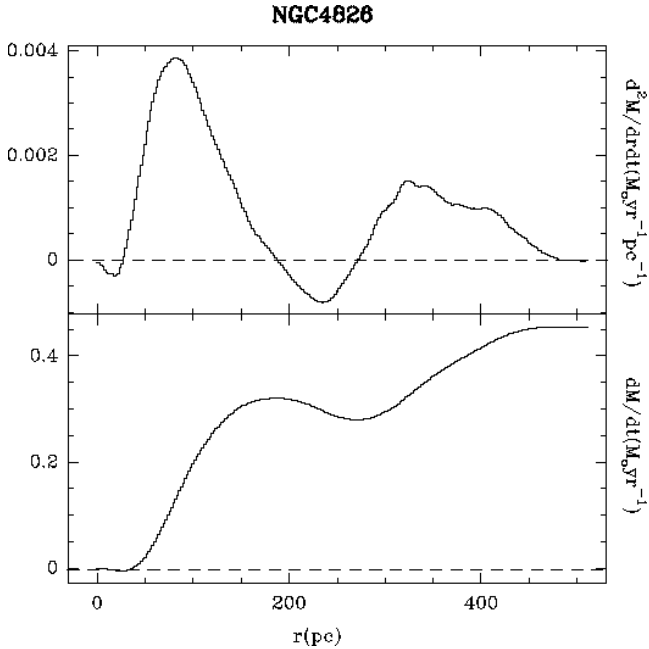}	      
   \caption{Same as Fig.~\ref{fig:dm-4321}\textit{a, b} but  for NGC\,4826. As is the case of 
the other transition object, NGC~4321, the overall budget in NGC\,4826 is clearly positive at all radii.
   }      
        \label{fig:dm-4826}
 \end{figure}

In spite of the large molecular gas concentration close to the AGN (r$<$150~pc), the overall feeding budget in 
NGC\,4826, quantified by the mass inflow rates shown in Fig.~\ref{fig:dm-4826}, is clearly {\it positive} inside 
the full field-of-view of the NIR image. This implies that the predicted radial mass flow 
should be dominated by outflow rather than inflow motions. 
At r$\sim$100~pc, dM/dt$\sim$+0.2M$_{\sun}$~yr$^{-1}$ and the inflow rate is still {\it positive} at
r$\sim$500~pc, where dM/dt$\sim$+0.45M$_{\sun}$~yr$^{-1}$.

\subsubsection{NGC\,4579\label{effi-n4579}}

Fig.~\ref{fig:pot-4579}\textit{a} shows the central r$\sim$1~kpc region of the I-band HST image of NGC~4579 
deprojected onto the plane of the galaxy. The dominant stellar perturbation in the HST image of the disk is the 
large-scale 9~kpc bar of NGC~4579, denoted BAR and oriented as shown in Fig.~\ref{fig:pot-4579}\textit{a}. As 
illustrated by Fig.~\ref{fig:pot-4579}\textit{b}, the potential strength of the bar is dominated by an $m=1$ mode 
for r$<$250~pc where Q$_{T}\geq$0.10.

Fig.~\ref{fig:torques-4579}\textit{a} shows the 2D pattern of gravitational torques for r$\leq$1~kpc. Stellar 
torques change sign as expected if the orientation of quadrants I-to-IV can be attributed to the 
large-scale stellar bar of NGC\,4579. Assuming that the sense of gas circulation is clockwise in
the galactic plane, Fig.~\ref{fig:torques-4579}\textit{a} shows that the bulk of the CO emission
along the spiral arms lies where torques are strong and negative.  
Fig~\ref{fig:torques-4579}\textit{b} shows that the azimuthally averaged torques are strong and negative from the edge 
of the image at r=1200~pc (and most probably from larger radii) down to r=200~pc. 

Closer to the AGN (r$<$200~pc), the bulk of the CO emission is concentrated North of the nucleus in the trailing 
quadrant I(+), where stellar torques are positive. The CO complex closest to the AGN
is in the leading quadrant II(-), East of the nucleus at r$\sim$150\,pc, where torques are negative. 
However, Fig.~\ref{fig:torques-4579}\textit{b} shows that the azimuthally averaged torques 
are positive for r$<$200~pc. As discussed in Sect.~\ref{obs-n4579}, the interpretation of 
the complex kinematics observed for molecular gas in this region is not straightforward 
but does not favor a direct feeding of the AGN. One of the possible scenarios explored
 by Garc\'{\i}a-Burillo et al.~2005 (in prep.) invokes that gas is being entrained by 
expanding motions at r$<$200~pc. Note that in this case we do not expect the gas pattern 
to be stationary with respect to the stellar pattern and therefore, the torque calculation 
developed above may not be relevant on scales r$<$200~pc. Nonetheless this would not change the picture 
where the central engine of NGC~4579 is not being fueled at present.

\begin{figure}[tb!]
   \centering
   \includegraphics[width=8.5cm]{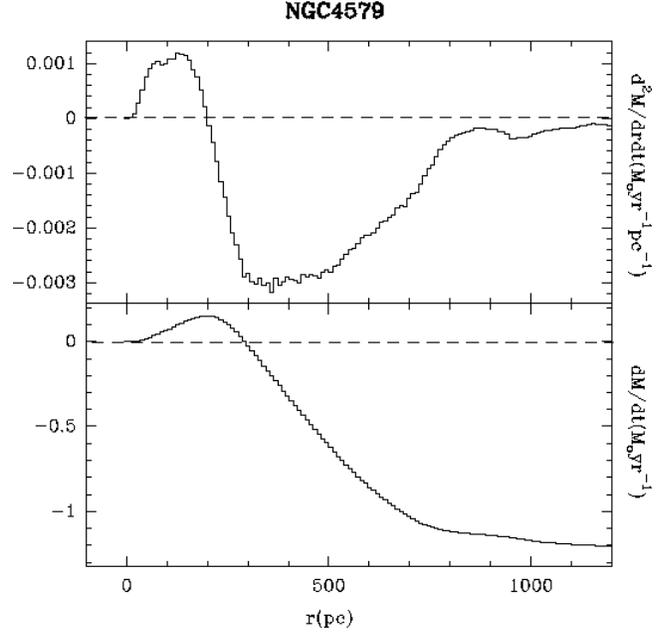}
   \caption{Same as Fig.~\ref{fig:dm-4321}\textit{a, b} but for NGC\,4579. The overall mass 
inflow budget is clearly negative down to r=300~pc due to the action of the large-scale bar. Inside this radius, 
stellar torques do not favour AGN feeding in this LINER/Seyfert.  
             }
         \label{fig:dm-4579}
 \end{figure}

In summary, the overall mass inflow budget is clearly negative down to r$\sim$200~pc due to the action of the large-scale bar 
of NGC~4579 (Fig.~\ref{fig:dm-4579}). The radial flow expected for the gas  
is clearly inward down to r$\sim$200~pc.  At r$\sim$500~pc, dM/dt$\sim$--0.6M$_{\sun}$~yr$^{-1}$. 
Inside r$\sim$200~pc stellar torques do not currently help feeding.  

\subsubsection{NGC\,6951\label{effi-n6951}}
 
Fig.~\ref{fig:pot-6951}\textit{a} shows the deprojected J-band HST image of the nuclear region (r$\leq$900~pc) of 
NGC~6951. The HST image allows the identification of the central 1.8~kpc of the large-scale bar of NGC\,6951 
(denoted as BAR in Fig.~\ref{fig:pot-6951}\textit{a}). The bar clearly extends beyond the image field-of-view 
(r=900~pc). Moreover we report on the detection of a small oval distortion most prominent 
inside r=400~pc (denoted as OVAL(n) in Fig.~\ref{fig:pot-6951}\textit{a}). The $m=2$ mode of the oval perturbation 
dominates the strength of the potential at r$<$400~pc. While it is true that we likely underestimate the contribution of the large-scale bar to the strength of the $m=2$ mode at r$<$400~pc due to the restricted field-of-view of the HST image, we do not expect this contribution to be significant on these scales, considering the low Q$_2$ values typically shown by large-scale bars at radii of a few 100~pc. The measured strength of the oval is moderate--to--low: Q$_T\sim$0.06, i.e., quite comparable to the strength of the oval potential detected in the nucleus of NGC~4826. We note that the strength of the large-scale bar is likely much higher than that of the oval but it is relevant only for the gas which lies beyond the r$\sim$900~pc circumnuclear CO disk.

The NGC~6951 torque map shown in Fig.~\ref{fig:torques-6951}\textit{a} suggests indeed that it is the nuclear 
oval which shapes the NGC~6951 torques map. The orientation of quadrants I-to-IV associated with the nuclear oval 
fits the pattern of the torques much better
than that associated to the bar. As the sense of gas circulation is clockwise in NGC~6951, 
Fig.~\ref{fig:torques-6951}\textit{a} shows that CO emission over the spiral arms in quadrants I(-) and III(-) 
lies at the leading edges of the nuclear oval where torques are negative down to the inner radius of the 
pseudo-ring, at r$\sim$300~pc. While the torques due to the large-scale bar on the inner CO ring are weak at 
present, torques are expected to be strong and negative at the leading edges of the bar. The offset dust lanes 
detected in the optical images of NGC~6951 (e.g., P\'erez et al.~\cite{per00}) indicate that gas inflow is at 
work along the bar on large scales. 
The combination of the bar and the nuclear oval has created a pattern of torques which are negative
down to r$\sim$300~pc. The bulk of the molecular gas in the central 5~kpc of NGC\,6951 
has already fallen from the x$_1$ orbits down to the x$_2$ orbits of the bar where it feeds a
nuclear starburst in the ring. In the vicinity of the AGN, however, the average stellar torques
become marginally positive and AGN feeding is not favored by the present configuration of the
stellar potential.

\begin{figure}[tb!]
   \centering
   \includegraphics[width=8.5cm]{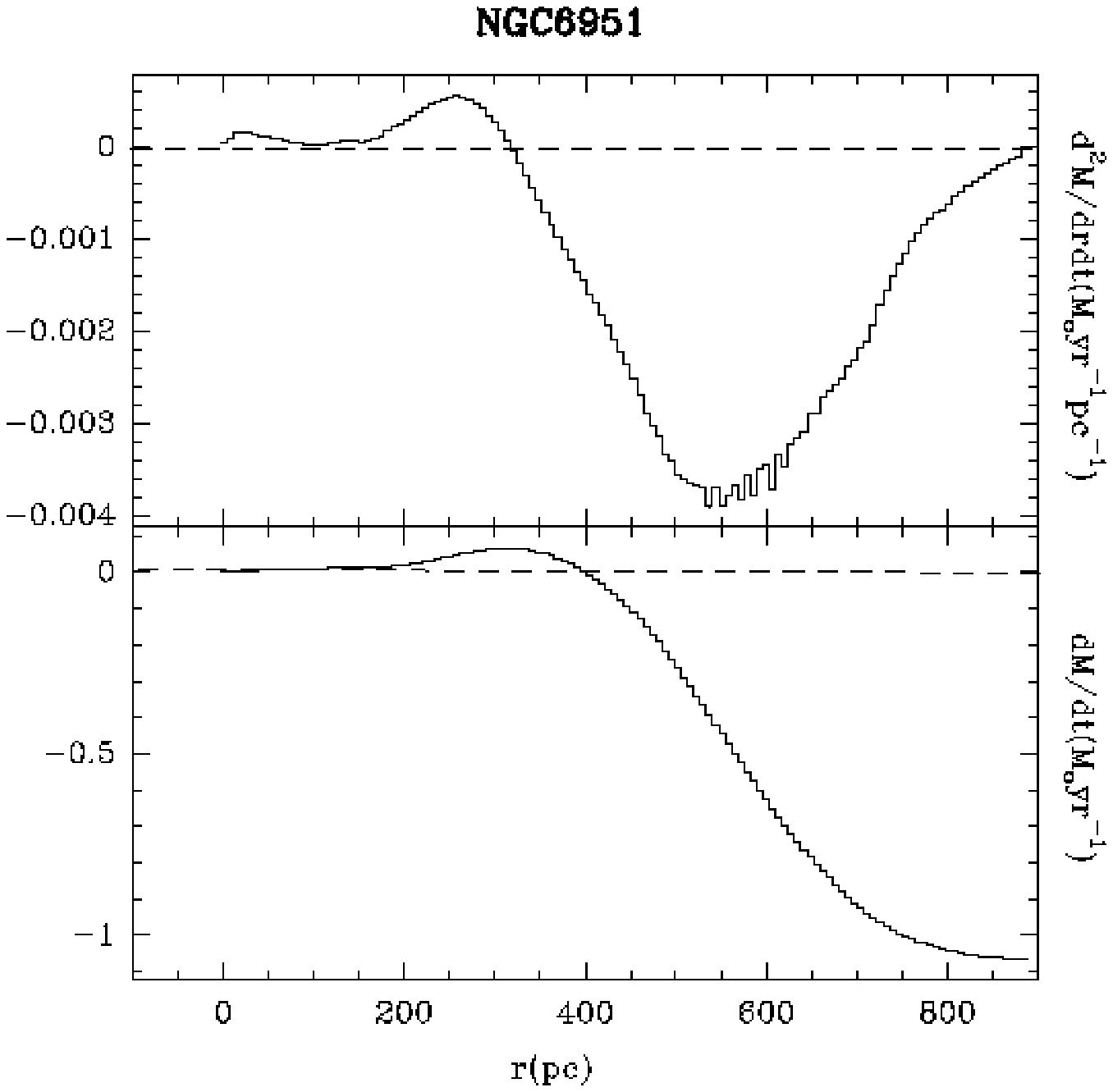}
   \caption{Same as Fig.~\ref{fig:dm-4321}\textit{a, b} but  for NGC\,6951. The overall mass 
inflow budget is clearly negative down to r=300~pc due to the combined action of the large-scale bar and the 
nuclear oval. Inside this radius, stellar torques do not favour AGN feeding in this Seyfert.
              }
         \label{fig:dm-6951}
 \end{figure}

Fig.~\ref{fig:dm-6951} shows the overall mass inflow budget for NGC~6951. The radial flow expected for the 
gas is clearly inward down to r$\sim$300~pc. At r$\sim$500~pc, dM/dt$\sim$--0.25M$_{\sun}$~yr$^{-1}$. 
Inside r$\sim$300~pc, however, stellar torques do not favour AGN feeding at present in this Seyfert.


\section{Discussion and conclusions\label{summary}}

The first results obtained from the analysis of stellar torques in the circumnuclear disks of 
a subset of NUGA targets reveal complex trends in the details of the feeding history along the
Transition/LINER/Seyfert sequence. The overall picture emerging in the two Seyferts/LINERs (i.e., NGC~6951
and NGC~4579) suggests that AGN feeding may proceed in two steps. In a first step, gravity torques
do help, by getting rid of gas angular momentum, driving the gas inwards and feeding a nuclear
starburst on scales of $\sim$a few 100~pc. On smaller scales, however, stellar torques play no role
in AGN fueling in the current epoch: torques on the gas are not negative all the way to the center, but on
the contrary they become positive and quench the feeding. This may explain why molecular gas seems
to 'avoid' the inner 200-300~pc of NGC~6951 and NGC~4579 where we measured
M$_{gas}<$a few$\times$10$^{5}$--10$^{6}$M$_{\sun}$ (see also the properties of NGC~7217 as discussed 
by Combes et al.~\cite{com04}). Observational evidence indicates that in
NGC~4579 molecular gas on these scales is presently flowing outward. In a second step, a mechanism 
complementary to gravity torques may be required to drive gas inflow on smaller scales.

Most notably, a much larger molecular gas concentration has been driven inwards in the transition
objects analyzed in this paper: $\sim$3$\times$10$^{7}$--10$^{8}$M$_{\sun}$ inside r=150~pc.
However, we have found that stellar torques are also unable to drain angular momentum from the
massive circumnuclear gas disks of NGC~4321 and NGC~4826. The underlying mechanism responsible for halting gas inflow inside r=150~pc
seems to be different in these two transition objects. On similar spatial scales, the stellar torques created by
the nuclear bar of NGC~4321 are much stronger than those created by the weak oval perturbation of
NGC~4826. Furthermore, the $m=1$ perturbation identified in the gas disk of NGC\,4826, which is probably unrelated to the oval perturbation, could be the main donor of angular momentum to the gas at these radii (paper I).

Our quest for a mechanism of AGN feeding based only on stellar torques has led to a
paradoxical result for the four cases studied here: inflow is actually thwarted close to the AGNs (r$<$100-200~pc).
One possible explanation is that the responsible agent in the stellar potential could be transient,
or as short lived as an individual AGN episode (hence: $\leq$10$^{7-8}$ years). The feeding phase
from 100~pc to 1~pc could be so short that the smoking gun evidence in the
potential is likely to be missed. Alternatively, this {\it temporary inability} of gravitational torques
could be overcome by other mechanisms that, over time, become competitive with non-axisymmetric
perturbations. Among the different mechanisms usually cited in the literature, dynamical friction
and viscous torques have been invoked to help AGN feeding on $\sim$100~pc scales. Gravitational
torques can become positive close to the central engine as illustrated above, in contrast
with viscous torques or dynamical friction that always favour gas inflow. This implies that stellar
torques could regulate the gas flows in galactic nuclei in combination with other
mechanisms that secularly drain the gas angular momentum.

We discuss below the efficiency of dynamical friction and viscosity versus gravity torques for
driving AGN feeding in the four galaxies analyzed in this paper. This efficiency can be quantitatively
estimated by comparing the typical time-scales of these processes with those of the 
gravitational torques derived above for each galaxy.

\subsection{Assisting gravity torques} 

\subsubsection{Dynamical friction}

Dynamical friction of giant molecular clouds (GMCs) in the stellar bulge of a galaxy is often
invoked as a possible mechanism of fueling AGN. According to the well-known Chandrasekhar (1943)
formula, the time-scale of dynamical friction for a GMC of mass M$_{GMC}$ at a radius $r$
where rotational velocity is $V_{rot}$ would be T$_{df}\sim$10$^8$ yr
(r/100~pc)$^2\times$(10$^6$ M$_\odot$/M$_{GMC}$)$^{-1}\times$($V_{rot}$/200kms$^{-1}$). 
We have derived the decay time-scale (T$_{df}$) of a typical GMC (M$_{GMC}\sim$10$^6$M$_\odot$) in
the nuclear disks of the 4 galaxies discussed above. We derive T$_{df}$ at the radii where we have
identified a barrier of positive gravity torques and get $V_{rot}$ from the CO observations.
The values of T$_{df}$ are typically $\sim$30, except in NGC\,4826 where we obtain formally T$_{df}
\sim$5 (see Table~\ref{time-scales}). The comparison between T$_{df}$ and the corresponding time-scales 
for gravity torques T$_{grav}$ (defined as T$_{grav}\sim L/\Delta L$ from Eq.~[\ref{tgrav}]) implies that 
friction is a relatively slow process compared to gravity torques, except in NGC\,4826 and NGC\,6951.

Despite these results, we suspect that the values estimated for T$_{df}$ are a strict lower limit and so
severely overestimate the efficiency of dynamical friction in all cases. The reason for this assessment is that
Chandrasekhar's formula implicitly assumes that a GMC is a bound rigid body, with a highly
concentrated mass. This approximation underestimates the real value of T$_{df}$. Molecular clouds
are indeed a loose ensemble of clumps of different sizes, resembling fractals. The typical masses of the
densest fragments can be as low as 10$^{-3}$ M$_{\odot}$ (e.g., Pfenniger \& Combes~\cite{pfe94}).
Therefore the scattering efficiency of a realistic GMC unit of 10--30~pc size including a complex
mass spectrum of clumps inside is much less than estimated above: the Coulomb parameter of a GMC in
Chandrasekhar's formula should tend to vanish. We therefore conclude that dynamical friction is
likely to be quite a slow process, which to first approximation can be neglected relative to 
gravity torques.

\begin{table}

\begin{tabular}{lccc}
\hline\hline
\multicolumn{1}{l}{Galaxy} &
\multicolumn{1}{c}{T$_{grav}$} &
\multicolumn{1}{c}{T$_{df}$} &
\multicolumn{1}{c}{T$_{visc}$}  \\
\multicolumn{1}{c}{} &
\multicolumn{1}{c}{(T$_{rot}$)} &
\multicolumn{1}{c}{(T$_{rot}$)} &
\multicolumn{1}{c}{(T$_{rot}$)}  \\
\hline\hline
{\bf NGC~4321}(r=200~pc) & 8 & 30 & 20  \\
{\bf NGC~4826}(r=50~pc) & 70 & 5 & 5   \\
{\bf NGC~4579}(r=200~pc) & 8 & 30 & 22   \\
{\bf NGC~6951}(r=200~pc) & 50--100 & 30 & 20    \\
\hline\hline
\end{tabular}
\caption{ 
We compare the characteristic time-scales of gravity torques
(T$_{grav}$), dynamical friction (T$_{df}$) and  viscosity (T$_{visc}$) in the nuclear disks of
NGC~4321 , NGC~4826, NGC~4579 and NGC~6951. Time-scales are given in units of 
rotation periods (T$_{rot}$).}
\label{time-scales}
\end{table}

\subsubsection{Viscosity}

Viscous torques are generally weak, and their associated time-scales (T$_{vis}$) are quite long at
large radii. The reason behind this is that T$_{vis}$ grows as the square of the radius according to the
classical diffusion equation of Pringle~(\cite{pri81}) which gives the time evolution of an
axisymmetric disk distribution, $\Sigma$($r$,$t$), under the action of viscous transport:





\begin{equation}
\displaystyle
\frac{\partial\Sigma}{\partial t} = -r^{-1} \frac{\partial}{\partial r} \displaystyle\Biggl[\frac{\partial [\nu \Sigma r^3(d\Omega/dr)]} {\partial r}\times \displaystyle\Biggl(\frac{d (\Omega r^2)}{dr}\displaystyle\Biggr)^{-1}\displaystyle\Biggr] 
\label{pringle}
\end{equation}

In this formula, the kinematic viscosity, $\nu$, is usually described as
$\lambda \times \sigma_v$, where $\lambda$ is the
characteristic mean-free path for viscous transport and $\sigma_v$ is the gas velocity dispersion.
Eq.~[\ref{pringle}] allows to derive by simple dimensional analysis the typical radial dependence of the time-scale for viscous transport 
in an axisymmetric disk, if we assume that the disk rotation curve can be described by a power law 
(i.e., V$_{rot} \propto r^{1-\alpha}$): 

\begin{equation}
\displaystyle
T_{vis}\sim\frac{2-\alpha}{\alpha} \times \frac{r^2}{\nu}
\label{tvis}
\end{equation}

\begin{figure}
\centering
\includegraphics[width=7.5cm]{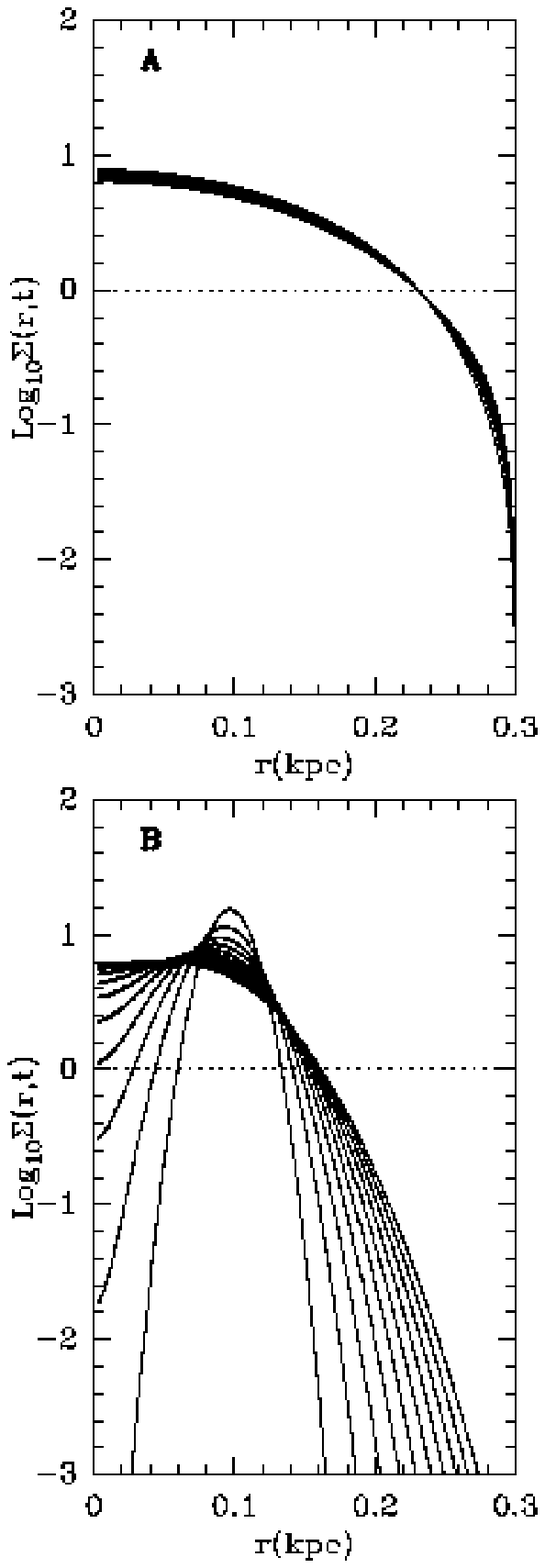}
\caption{
{\bf a)} The time evolution of gas surface density in an axisymmetric disk,
where the rotation curve is of the form: V$_{rot} \propto r^{1-\alpha}$, with
$\alpha=0.1$. To normalize, we fit V$_{rot}$=200~km~s$^{-1}$ at $r$=1~kpc. Initially the gas is
distributed according to a surface density law proportional
to $\Sigma \propto 1+ cos (\pi r/r_c)$, until the cut-off radius
$r_c$ = 0.3~kpc. We represent Log$_{10}\Sigma$(r,t)
in arbitrary units. The various curves are the evolution
for times spaced by 2$\times$10$^{7}$ yr, until 2$\times$10$^{8}$.
The column density profile shows hardly any change after 2$\times$10$^{8}$ yr.
{\bf b)} Same as {\bf a)} but here for an initial gas distribution in the shape a
narrow ring between 90 and 110~pc radii (top-hat profile). The
$\Sigma$(r,t) profile changes significantly already after 5$\times$10$^{7}$yr.}
\label{visco}
\end{figure}

However, as illustrated below, the exact value of $T_{vis}$ strongly depends on the assumed initial 
density distribution law for the disk. Fig.~\ref{visco}\textit{a} shows the time evolution of a smooth disk 
distribution under the action of viscous torques. In this calculation we have taken $\nu$=100~pc~km~s$^{-1}$, assuming
$\lambda$=10~pc and $\sigma_v$=10~km~s$^{-1}$.  After 200~Myr the
radial gas flows are very small. However, the efficiency of viscous transport can be enhanced if
the initial distribution of the disk is characterized by strong density gradients. This is
the case of nuclear contrasted rings, especially when these are located in the inner
regions of galaxies (r$\sim$100--500~pc) as frequently is the case for early-type
barred spirals. Moreover, and contrary to the conventional wisdom (based on low resolution
observations) that most galaxy rotation curves are close to rigid body for r$<$500~pc, $\Omega$ can
show a strong variation with radius in the inner regime. Therefore galactic shear can still be
very high on these scales; this also favours viscous transport.

Fig.~\ref{visco}\textit{b} illustrates the time evolution of a nuclear ring of
r$\sim$100~pc under the action of viscous torques in a galaxy disk. The radial gas flows
are remarkably faster than in the previous case: in typically 5$\times$10$^7$ yr, the ring starts to dissolve and a
significant amount of gas can reach the center of the galaxy. We compare in Table~\ref{time-scales} the characteristic time-scales of viscous torques and gravity torques in the 4 galaxies analyzed in this paper.  Power-law rotation curves have been fit within the 
relevant radial ranges adapted to each galaxy, assuming in all cases $\nu$=100~pc~km~s$^{-1}$, i.e., a viscosity prescription 
similar to the the axisymmetric smooth disk solution. Viscosity seems a viable mechanism to drive significant gas inflow in NGC\,4826 at r=50-100~pc, while it is probably inefficient against the stronger gravity torques in NGC\,4579 and NGC\,4321 for r$<$200~pc.  Moreover, gas lying inside the nuclear ring of NGC\,6951 could be falling to the AGN favoured by viscous torques at r$<$200~pc.
Note that these results are unchanged if we allow gravity torques to be a factor of 1.5--2 stronger,
as argued in Sect.~\ref{method}.

We can conclude that, in spite of all the uncertainties in the derivation of time-scales, viscous
torques are the most viable mechanism to generate gas inflow on scales $\sim$100-200~pc if they act
on a contrasted nuclear ring distribution, and do not have to fight against very strong positive
torques from gravity.

\subsection{A scenario for self-regulated activity in LLAGNs}
In this work we have shown that gravity torques exerted by the 
stellar potential on the gas disks of four LLAGNs, purposely chosen to 
represent the whole range of activity classes in the NUGA sample, fail 
to account for the feeding of the AGN at present. We conclude that 
gravity torques need to be assisted in due time to drive the gas to 
the center. Based on these results, we develop in this section a 
simplified general scenario in which the onset of nuclear activity 
can be understood as a recurrent phase during the typical lifetime of 
any galaxy. In this scenario the recurrence of activity in galaxies is indirectly related to that of the bar
instabilities although, as argued below, the active phases are not necessarily coincident with the
maximum strength of a single bar episode. While bars can build up gas reservoirs towards the
central regions of galaxies in the shape of nuclear rings in early type objects, the dynamical
feed-back associated with these gas flows can destroy or considerably weaken the bars 
(Norman et al.~\cite{nor96}; Bournaud \& Combes~\cite{bou02}). When this happens, gravity torques are negligible and can make 
way for other competing mechanisms of gas transport, such as viscous torques, in an almost axisymmetric system.

 
The first step in this evolutionary scenario could begin with
an axisymmetric disk of gas. Shortly after the galaxy disk would be prone to a bar instability. If
the galaxy is early-type, i.e., with a sufficient bulge or central mass concentration,
then the bar pattern is likely to have one or even two Inner Lindblad Resonances (ILRs). A nuclear
ring would form either close to the single ILR or between the two ILRs. At this stage, the gas
would be driven inwards from corotation to the ILR, and would accumulate in the ring where a burst
of star formation can occur. However positive gravity torques would prevent the gas from flowing further
in: under the bar forcing, gas inside the ring could be evacuated outwards (Combes~\cite{com88}). The bar thus
would restrict the amount of fuel available for AGN feeding to that lying very close to the center
where it is under the dominant gravitational influence of the black hole. As the potential would be 
mostly axisymmetric in the central $\sim$10~pc, gas could remain there feeding the central engine.

The infall of gas driven by a bar is self-destructive, however: it progressively weakens and
destroys the bar, so that the potential returns to axisymmetry (Norman et al.~\cite{nor96}; Bournaud \&
Combes~\cite{bou02}). At this stage gas piled up in the
nuclear ring could dissolve to form a smoother disk through viscosity which may have become
competitive against gravity torques. The infall of gas produced by viscous torques could replenish
the exhausted fuel near the central engine triggering a new phase of nuclear activity.

In this scenario, the feeding of active nuclei in early-type spirals is contemplated as a two step process:
first gravity torques bring the gas of the large-scale disk to the nuclear ring, and when the
dynamical feedback has destroyed the bar, the viscous torques could smooth out the ring, and bring
gas into the central 10~pc, where it is under the influence of the Keplerian potential of the black
hole. The disk becomes axisymmetric and the cycle can be restarted at its first step. In
particular, the disk will be prone to a new bar instability if gas is accreted from the outer parts
of the disk. These multiple bar phases must occur in gas-rich spiral galaxies, in order to account
for the observed frequency of bars (Block et al.~\cite{blo02}; Bournaud \&
Combes~\cite{bou02,bou04}).

We relate the recurrence of episodes of bar formation and destruction with the fueling of active nuclei in galaxies. The self-regulated competition between gravity torques and viscosity in galaxy nuclei may lead typically 
to several episodes of activity (each lasting for $\sim$ 10$^{8}$~yr) during a single cycle of bar formation/destruction 
(lasting for 10$^{9}$~yr). These activity episodes are not expected to be  strongly correlated with the phases of maximum strength for the bar, but they may appear at different evolutionary stages of the bar potential, depending on the balance between gravity torques and viscosity. The observed prevalence of outer rings in Seyferts, interpreted as a sign of bar dissolution, would  support this scheme (Hunt and Malkan~\cite{hun99}). Furthermore, the ample variety of 
morphologies revealed by the CO maps of the circumnuclear disks of NUGA targets corroborates that there is  no universal pattern associated with  LLAGNs (Garc\'{\i}a-Burillo et al.~\cite{gb04}). Activity can also be found in a galaxy during its 
axisymmetric phase (e.g., see the case of the NUGA galaxy NGC~5953, discussed by Combes et al.~2005, in prep).

The gas directly responsible for the Seyfert activity in galaxies like
NGC~4579 or NGC~6951 must have been brought to the center during the previous
axisymmetric phase, while the bar at present is emptying the region inside the nuclear 
ring, thus regulating the amount of gas available for the active nucleus. In the case of NGC~4579,
gravity torques clearly overcome viscosity, while in NGC~6951 the balance between both mechanisms
might be reached soon. In the case of NGC~4826, the nucleus has a very low level activity in spite
of the presence of a large amount of gas in the vicinity, since the potential is almost
axisymmetric, and the nuclear ring could be undergoing dissolution by viscosity. The active phase would
develop during the next phase, while the disk would soon become bar unstable again. However, the
role of the $m=1$ gas perturbations identified in the disk of NGC\,4826 might be to slow down the
dissolution of the ring, as kinematics still suggest outward motions close to the AGN. The case
of NGC~4321 is still early in the time evolution, as the bar is only now entering the destruction
phase, through the formation of a nuclear bar inside the resonant ring. This nuclear bar efficiently
prevents nuclear feeding through gravity torques.

The observed fraction of LLAGNs in the Local Universe ($\sim$44$\%$, including LINERs; Ho et al.~\cite{ho97}) could impose tight constraints on the expected number of activity episodes per bar formation/destruction cycle in this scenario. Based on these statistics, a rough estimate would raise this number to $\sim$4. However this estimate can only be taken as mostly speculative for the time being.  The principal source of uncertainty, which prevents us from making a more quantitative prediction, resides in the fact that there is no consensus on what is the typical duration of a nominal AGN duty-cycle. The different values of AGN duty-cycles discussed in the literature range from 10$^{7}$ to 10$^{8}$~yr (Ho et al.~\cite{ho03}; Martini~\cite{mar04}; Wada~\cite{wad04}; Merloni~\cite{mer04}). Moreover, the required number of individual episodes could be lowered to $\sim$a few if we exclude starburst dominated LINERs from the LLAGN population or even more if we restrict the activity classification to Seyferts, the most active members in the family.

In barred galaxies with comparatively less prominent bulges, i.e., characterized by the
absence of an ILR barrier, viscosity may not be needed to take over gravity torques close to the
AGN. The gas flow should not be stalled at a radius of a few $\sim$100~pc forming a nuclear ring in
this case. There are a few examples in the NUGA survey of strongly barred galaxies showing a strong
nuclear concentration of gas instead of a ring. This class of ILR-free bars could be more
common in the early Universe during the formation of disks and the growth of supermassive black
holes (Sellwood \& Shen~\cite{sel04}).

Furthermore, other mechanisms different from axisymmetric viscous flow could
be efficient at overcoming the ILR barrier imposed by most stellar bars in the Local Universe. 
As mentioned above, the ultimate agent for the feeding could still reside in the stellar potential 
but would be transient and their detection might be elusive. In addition, the role of nuclear $m=1$ and $m=2$ gaseous
spirals in AGN feeding is still not completely elucidated. Most interestingly, a few examples of
these gas instabilities are found in the available CO maps of AGNs, including some of the galaxies
of the NUGA sample (Garc\'{\i}a-Burillo et al.~\cite{gb04}). 

Finally, the role of gas self-gravity to drive gas inflow may not be negligible in those cases where 
the distribution of gas is a significant source of non-axisymmetry for the 
total gravitational potential, especially if the stellar potential itself appears as featureless or mostly 
axisymmetric. We can estimate the influence of gas self-gravity by calculating the typical gas mass 
fractions as a function of radius in the case of the galaxies studied in this paper. 
In NGC~6951 and NGC~4579, the gas mass fractions are very low throughout the disk from r$\sim$100~pc ($\sim$1$\%$)
up to r$\sim$1000~pc ($\sim$3$\%$). Therefore we do not expect gas instabilities to have a 
significant influence in the total gravitational potential in these two galaxies. In contrast, gas mass fractions
are comparatively larger at the same radii in NGC~4826 and NGC~4321: $\sim$10-15$\%$ at r$\sim$100~pc in both galaxies.
This gas mass fraction is still high at r$\sim$1000~pc ($>$10$\%$) in NGC~4321, while it decreases down to $\sim$5$\%$ in NGC~4826
at that distance. This evidence points to a larger influence of gas self-gravity in NGC\,4826 (as discussed
in paper I) and NGC\,4321 (as discussed by Wada et al.~\cite{wad98}).

These results highlight the need for a 
significant increase in the number objects for which high-resolution CO maps are available together
with careful case-by-case studies in the quest for evidence of AGN feeding. 
The scenario proposed in this work, as well as any alternative model accounting for nuclear activity in galaxies, 
remains to be tested using a larger sample of LLAGNs.

\begin{acknowledgements}
	 We thank the anonymous referee whose comments helped to improve this paper. We would like to thank Andrew~J.~Baker for his careful reading of the paper and related discussions. We acknowledge the IRAM staff from the Plateau de Bure and from 
         Grenoble for carrying out the observations and help provided during the 
	 data reduction. This paper has been partially funded by the Spanish MCyT 
         under projects DGES/AYA2000-927, DGES/AYA2003-7584, ESP2001-4519-PE 
         and ESP2002-01693, and European FEDER funds.

\end{acknowledgements}

\end{document}